\documentclass[useAMS,usenatbib]{mn2e}
\usepackage{graphicx}
\usepackage{fixltx2e}
\usepackage{times}
\usepackage{amsmath}
\usepackage{amsfonts}
\usepackage{url}
\usepackage{placeins}
\usepackage{natbib}
\usepackage{amssymb,amsmath}
\usepackage{url}
\usepackage{placeins}

\newcommand{\mhbar}{{\bar{M}_h}}
\newcommand{\Mhmin}{{M_{h,\rm min}}}
\newcommand{\Mhmax}{{M_{h,\rm max}}}
\newcommand{\Lmin}{{L_{\rm min}}}
\newcommand{\Lmax}{{L_{\rm max}}}
\newcommand{\Mh}{{M_{\rm h}}}
\newcommand{\hmpc}{\,h^{-1}\rm Mpc}
\usepackage[normalem]{ulem}
\usepackage{soul}
\newcommand{\fduty}{f_{\rm duty}}

\begin{document}

\title[Clustering of intermediate redshift quasars]{Clustering of intermediate redshift quasars using the final \textsc{sdss iii-boss} sample}
\author[Eftekharzadeh et al.]{Sarah Eftekharzadeh$^{1}$, Adam D. Myers$^{1}$, Martin White$^{2,3}$, David H. Weinberg$^{4}$, \newauthor
Donald P. Schneider$^{5}$, Yue Shen$^{6,7}$, Andreu Font-Ribera$^{8}$, Nicholas P. Ross$^{9,10}$,\newauthor Isabelle Paris$^{11}$, Alina Streblyanska$^{12,13}$ \\
\\
$^{1}$ Department of Physics and Astronomy, University of Wyoming, 1000 University Ave., Laramie, WY, 82071, USA \\
$^{2}$ Physics Division, Lawrence Berkeley National Laboratory,
1 Cyclotron Rd., Berkeley, CA 94720, USA \\
$^{3}$ Departments of Physics and Astronomy, 601 Campbell Hall,
University of California Berkeley, CA 94720, USA \\
$^{4}$ Department of Astronomy and CCAPP, Ohio State University,
  Columbus, OH, USA \\
$^{5}$ Department of Astronomy \& Astrophysics and Institute for Gravitation \&
the Cosmos, Pennsylvania State University, University Park, PA 16802 \\
$^{6}$Carnegie Observatories, 813 Santa Barbara Street, Pasadena, CA 91101, USA \\
$^{7}$Kavli Institute for Astronomy and Astrophysics, Peking University, Beijing 100871, China \\
$^{8}$Lawrence Berkeley National Laboratory, 1 Cyclotron Road, Berkeley, CA, 94720 USA \\
$^{9}$Department of Physics, Drexel University, 3141 Chestnut Street, Philadelphia, PA 19104, USA \\
$^{10}$Institute for Astronomy, University of Edinburgh, Royal Observatory, Edinburgh EH9 3HJ, UK \\
$^{11}$INAF - Osservatorio Astronomico di Trieste, Via G. B. Tiepolo 11, I-34131 Trieste, IT\\
$^{12}$Instituto de Astrofisica de Canarias (IAC), E-38200 La Laguna, Tenerife, Spain \\
$^{13}$Departamento de Astrofisica, Universidad de La Laguna (ULL), E-38205 La Laguna, Tenerife, Spain}

\date{\today}
\maketitle

\begin{abstract}

We measure the two-point clustering of spectroscopically confirmed quasars from the final sample of the Baryon Oscillation Spectroscopic Survey (\textsc{boss}) on comoving scales of $4 \lesssim s \lesssim 22 ~  h^{-1} {\rm Mpc}$. The sample covers $6950\,\rm deg^{2}$ ($\sim 19\,(h^{-1}\rm Gpc)^{3}$) and, over the redshift range $2.2 \leq z \leq 2.8$, contains 55,826 homogeneously selected quasars, which is twice as many as in any similar work. We deduce $b_Q = {3.54\pm0.10}$; the most precise measurement of quasar bias to date at these redshifts. This corresponds to a host halo mass of $\sim 2\times10^{12} ~ h^{-1}M_\odot$ with an implied quasar duty cycle of $\sim 1$ percent. The real-space projected correlation function is well-fit by a power law of index −2 and correlation length $r_{0}=(8.12 \pm 
0.22)\,h^{-1}\rm Mpc$ over scales of $4 \lesssim r_{\rm p} \lesssim 25 ~ h^{-1} {\rm Mpc}$. To better study the evolution of quasar clustering at moderate redshift, we extend the redshift range of our study to $z\sim 3.4$ and
measure the bias and correlation length of three subsamples over $2.2 \leq z \leq 3.4$. We find no significant evolution of $r_{0}$ or bias over this range, implying that the host halo mass of quasars decreases somewhat with increasing redshift. We find quasar clustering remains similar over a decade in luminosity, contradicting a scenario in which quasar luminosity is monotonically related to halo mass at $z\approx 2.5$. Our results are broadly consistent with previous \textsc{boss} measurements, but they yield more precise constraints based upon a larger and more uniform data set.

\end{abstract}

\section{Introduction}
\label{intro}
It has long been known that a supermassive central black hole occupies
every massive galaxy \citep[e.g.,][]{kor95}, and that the galaxy and the black
hole properties are strongly correlated \citep[e.g.,][]{mag98,fe00,ge00}. 
Galaxies and their black holes therefore appear to have co-evolved.
 Although, there are other non-causal explanations that have been invoked to explain these relations \citep[e.g.,][]{koho13}.
This situation could arise because feedback from quasars partially regulates star
formation, and thus the properties of galaxy bulges, through merger-driven
winds, gas accretion or stochastic mergers \citep[e.g.,][]{nan07,sil08,Shan09}.
Alternatively, this relationship could occur because both supermassive black holes and their host galaxies
are governed by the properties of their parent dark matter halo
\citep[e.g.,][]{Fer02}.
Quasar populations show a high level of clustering, and it is inferred that
they occupy dark matter halos of $\sim 10^{12}\,h^{-1}M_\odot$ at most epochs
\citep[e.g.,][]{po04,cro05,my06,my07,she07,da08,ro09,wh12}.  Star formation is also most efficient in halos
of this mass \citep[e.g.,][]{mos10,beh10,bet12,vie13halomass}, providing further circumstantial
evidence for a link.
Further study of the relation between quasar activity and galaxy evolution
is clearly warranted.

The geometry of the cosmological world model is now known with remarkable
precision, and even systematic differences are at the few percent level
\citep[e.g.,][]{WMAP,Planck,and14,aub14}.
The spectrum of primordial fluctuations and its evolution over cosmic
history is also becoming increasingly well constrained.
Armed with this knowledge, it has become increasingly meaningful to study
galaxy formation over a range of redshifts---particularly at high redshift,
where subtle changes in the cosmological model can greatly affect 
predictions about galaxy evolution.
Driven by this better cosmological understanding, coupled with large
surveys of active galaxies over a range of wavelengths, supermassive
black holes are increasingly being fit into a wider cosmological framework
for galaxy formation \citep[see][for a comprehensive review]{alhic12}. 

{}From the quasar perspective, a wider understanding of the interplay
between galaxies, star formation and black holes has been studied extensively in extragalactic surveys
(again see \citealt{alhic12} as well as \citealt{Hop07,co13,Vea14,cap14}) 
. In addition to direct measurements 
of the relationship between the luminosity of Active Galactic Nuclei (henceforth AGN) luminosity and star formation in galaxies \citep[e.g.,][]{ 
alhic12,mul12,ros12}, great progress has been driven 
by two statistical measures of the quasar population: the luminosity function 
and clustering.
The quasar luminosity function is increasingly well-understood
\citep[e.g.,\ see][for a recent study]{ro13}
and there are now a large number of quasar clustering measurements in the
optical at $z < 2.2$ across a range of scales
\citep[e.g.,][]{po04,cro05,po06,he06a,my07,my07b,my08,ro09}. 
Together, these measurements provide an increasingly coherent characterization
of the dark matter environments inhabited by quasars and {\em what fraction}
of quasars ignite in those environments
\citep[i.e.,\ the quasar duty cycle;][]{col89,ha01,mw01,wh08,shan10a}.
Measurements of quasar clustering in the optical have informed
similar studies at $z < 2$ from samples selected in the X-rays
\citep[e.g.,][]{All11,Kru12} in the radio \citep[e.g.,][]{she09a},
in the infrared \citep[e.g.,][]{Don14,Dip14,Dip14CMB,hic11}
and, in general, in AGN samples selected across the electromagnetic
spectrum \citep[e.g.,][]{hi09}.
Beyond $z < 2$ relatively weak direct constraints on quasar clustering
at $z > 2.9$ also exist \citep{she07,She10} and similar high-redshift
measurements are starting to be made for quasar samples identified outside
of the optical \citep[e.g.,][]{All14}.

It is now critical to understand quasar clustering in the range $2<z<3$
(or at ``moderate redshift'' for quasars) for a number of reasons.
Luminous quasars peak in number density at moderate redshift
\citep[e.g.,][]{ro13},
so this is the most promising epoch for understanding how quasar feedback
and dark matter environment are related.
Measurements of quasar clustering are generally sparse at moderate redshift,
and such redshifts may be key to linking the relatively weak constraints
at $z > 3$ to the more accurate measurements at $z < 2$. Most notably, \citet{she07}
found (for a power-law index of $\gamma=2.0$) that quasars cluster with $r_{0}=16.9\,h^{-1}\rm Mpc$
at $z\sim3$ growing to $r_{0}=24.3\,h^{-1}\rm Mpc$ at $z > 3.5$. It remains
unclear exactly how these large correlation lengths transition to $r_{0}\sim5\,h^{-1}\rm Mpc$
($\gamma\sim2.0$) at $z < 2$  \citep[e.g.,][]{po06}.
In addition, quasar clustering measurements in samples selected beyond the optical are
now being made at moderate redshift, and comparisons between these samples
and (rest-frame) UV-luminous quasars have historically been critical to
interpret the quasar phenomenon at $z < 2$.
Star formation has been shown to occur most efficiently in halos of similar
mass to those occupied by quasars at
$z < 2$ ($\sim 10^{12}\,h^{-1} \rm M_{\odot}$), and as measurements of star formation as a function of dark matter halo
mass are now being pushed to $z > 2$ \citep[e.g.,][]{vie13,vie14} it will
be intriguing to see whether this trend continues to higher redshift.
Finally, models of how quasars populate dark matter halos, for instance,
empirical approaches such as the halo occupation distribution framework
\citep[HOD;][and references therein]{sel00,pea00,ben00,sco01,whi01,bw02,coo02}
have been successful in modeling galaxies at low redshift
\citep[e.g.,][]{zh05,zh07,Zeh11},
but are now being updated to model quasar clustering at different wavelengths
and redshifts \citep[e.g.,][]{Deg11,ch12,ric12,wh12,co13,Ric13,Vea14,cap14}.
It is not clear to what extent such HOD models describe rare and
heavily biased quasars at high redshift, and it is critical to push
observations to $z\sim3$ as the most sophisticated hydrodynamic simulations
being used to interpret the HOD are now being pushed from high redshift down
to $z < 4$ \citep[e.g.,][]{Deg12,Dim12}.

Given the importance of new observational constraints 
on quasar clustering at moderate redshift, we have embarked on 
measurements of quasar clustering at $2 < z < 3$ using a set of quasars from 
the Baryon Oscillation Spectroscopic Survey \citep[\textsc{boss};][]{Daw13} part of
the third phase of the Sloan Digital Sky Survey \citep[\textsc{sdss-iii};][]{ei11}. Although the
main goal of {\sc boss} was to constrain cosmological models, such a
large sample of quasars offers the chance to study galaxy evolution through measurements
of quasar clustering in the largely unsampled redshift range of $2.2 < z < 2.8$.
{\sc boss}  reaches more than two magnitudes deeper than the original \textsc{sdss} spectroscopic survey \citep[]{rich02a,sch10}. This increased depth, combined with improved targeting algorithms, allowed BOSS to spectroscopically confirm 15 times as many $z\sim2.5$ quasars as earlier iterations of the SDSS.

\citet{wh12} reported preliminary quasar clustering measurements from {\sc boss} using the autocorrelation of $\sim
27{,}000$ quasars spread
over $\sim 3600$\,deg$^2$. \citet{wh12} found a quasar bias of $b_Q = {3.8\pm0.3}$ at $\bar{z}\sim2.4$. 
Using a similar {\sc boss} sample, Font-Ribera et al. (2013) cross-correlated $\sim60{,}000$ quasars with the 
Ly$\alpha$ forest and derived $b_Q =3.64^{+0.13}_{-0.15}$ at  $\bar{z}\sim2.4$. Font-Ribera et al.\  could use
a larger sample size because their adopted cross-correlation technique is insensitive to the angular
selection function of {\sc boss} quasars. In this paper, we update the \citet{wh12}
 {\sc boss} quasar autocorrelation results
using the considerably larger and final {\sc boss} sample. {\sc boss} has now
spectroscopically confirmed 
$\sim160{,}000$ $z > 2.2$ quasars to a depth of $g < 22$ (and/or $r < 21.85$).
About half of these quasars are uniformly-selected, with
about two-thirds of that uniform, or ``\textsc{core}'' sample populating a large
contiguous area of
$\sim7500$\,deg$^2$ in the North Galactic Cap \citep[\textsc{ngc};][]{ros12}.
This $\sim7500$\,deg$^2$ is by far the largest
contiguous area ever used for $z > 2.2$ quasar clustering measurements, and
represents a contiguous area $\sim3\times$
larger than used in \citet{wh12}.
The main {\sc boss} quasar sample and different subsamples that we use for our
measurements are described in $\S $\ref{dat}. The clustering method and
uncertainty estimation are presented in $\S $\ref{cls} and the clustering
results are presented in $\S $\ref{clsres}.
The evolution of real and redshift-space correlation functions in terms of
redshift and luminosity are discussed in $\S $\ref{ev}. 
In \S\ref{sec:HM} we interpret our measurements in terms of characteristic
halo masses and associated duty cycles, with particular attention to the
observed (lack of) luminosity dependent clustering.
We summarize our conclusions in \S\ref{sec:conclusions}.
Appendix A introduces the software that we used
to generate a uniform random catalog from the survey mask.
We adopt a $\Lambda$CDM cosmological model with $\Omega_{m}=0.274$,
$\Omega_{\Lambda}=0.726$, $h=0.7$ and $\sigma_{8} = 0.8$ as assumed in other
{\sc boss} analyses \citep[e.g.,][]{wh11,re12,wh12,an12}. We express magnitudes
in the AB system \citep{Oke83}.

\begin{figure*}
\begin{center}
\includegraphics[angle=270,scale=0.50]{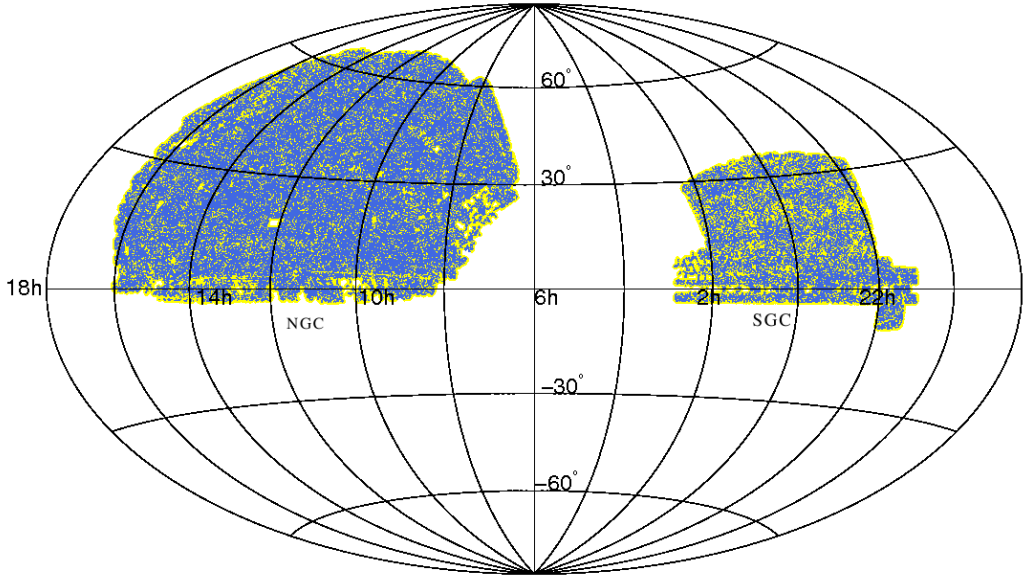}
\caption{Aitoff projection of the angular distribution of quasars in the final
{\sc boss} sample in equatorial coordinates. The background (yellow) depicts all sectors of the
\textsc{boss} survey, as well as a random catalog (for all survey sectors)
produced using the \textsc{bossqsomask} software (see the Appendix).The (blue)
dots are \textsc{boss} \textsc{core} quasars in survey sectors that have a
completeness greater than 75\%. The \textsc{core} sample corresponds to
objects that have been uniformly-selected using the {\sc xdqso} algorithm
\citep{bo11}.} 
\label{aitoff}
\end{center}
\end{figure*}

\section{Data} 
\label{dat}

\subsection{BOSS Quasars}

The \textsc{sdss-i/ii/iii} imaging surveys mapped over 14{,}000 $\rm {deg^{2}}$ of sky using the 
\textsc{sdss} camera \citep{gu98} mounted on a 2.5-meter telescope \citep{gu06} located at the Apache Point Observatory.
Of the unique imaging area released as part of \textsc{sdss} DR8 \citep{DR8},
the {\sc boss} survey targeted quasars over $\sim 7400\,{\rm deg}^{2}$ of a
region around the North Galactic Cap (\textsc{ngc}) and $\sim 2600\,{\rm
deg}^{2}$ of a region around the South Galactic Cap (\textsc{sgc}).
The Baryon Oscillation Spectroscopic Survey \citep[{\sc boss};][]{Daw13}, one of
four \textsc{sdss-iii} surveys \citep{ei11}, has the primary goal of measuring
the clustering of luminous red galaxies and neutral hydrogen to determine the
cosmic distance
scale\footnote{https://www.sdss3.org/surveys/boss.php}.
As part of this effort, \textsc{boss} measured spectroscopic redshifts of $1.5$ million galaxies and $190{,}000$
quasars using twin multi-object fiber spectrographs \citep[]{sm13}.
In this paper, we focus on using these data to measure quasar clustering near $z\sim2.5$.

{\sc boss} quasars are selected from point sources in 5-band $ugriz$ imaging \citep[]{fu96} that
meet a selection of photometric flag cuts \citep[see the appendixes of][]{bo11,ros12} 
and that fall within magnitude limits of $i > 17.8$ and dereddened $ g \le 22.0$ or $ r \le 21.85$.
Targeting quasars near $z\sim 2.5$ is 
difficult, as metal-poor stars typically have colours very close to those of
unobscured quasars around redshift 2.7 \citep[e.g.,][]{fan99, rich01}, and
quasars at redshifts of 2.2 to 2.6 can have similar colours to quasars at
redshift of about 0.5 because host galaxy light mimics the Lyman-$\alpha$ Forest
entering the $u$-band \citep[e.g.,][]{bu01,rich01,wei04}. These colour
similarities are
exacerbated near the faint limits of imaging, where flux uncertainties are larger. To circumvent
these issues, \citet{bo11} developed the {\sc xdqso} algorithm to target quasars for {\sc boss}
spectroscopic observation.  {\sc boss} targeted all sources with a ($2.2 < z < 3.5$) {\sc xdqso}
probability of {\tt pqsomidz} $> 0.424$ tuned to produce 20\,{\rm deg}$^{-2}$ quasar targets
over the {\sc boss} footprint \citep{ros12}.  This was known as the ``\textsc{core}'' sample.

The primary mission of the {\sc boss} quasar survey was to study quasar clustering in the Lyman-$\alpha$ Forest
via methods that are insensitive to the quasar selection technique.
As such, a heterogeneous range of multi-wavelength data was used to target {\sc boss} quasars,
and a  number of methods were incorporated into the survey in addition to {\sc xdqso} 
(\citealt{Ric09,Yec10,Kir11}; see \citealt{ros12}).
This heterogeneous or ``\textsc{bonus}'' approach augmented the \textsc{core} sample.

In this paper we focus on {\sc boss} quasars spectroscopically confirmed prior to 5th April 2014,
which is essentially the final {\sc boss} quasar compilation. These quasars
correspond to {\sc boss} spectral reduction pipeline version 5-7-0
(\texttt{spall-v5\_7\_0}). Figure \ref{aitoff} shows the
angular distribution of our quasar sample. The yellow background shown in Figure \ref{aitoff} represents the survey area,
and the point sources are the \textsc{core} quasars over the redshift range $2.2
\le z \le 2.8$.

\begin{figure}
\begin{center}
\includegraphics[angle=0,scale=0.30]{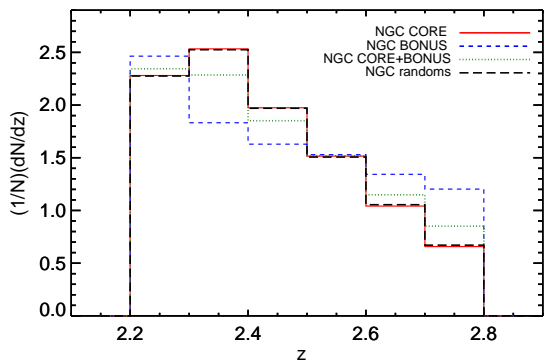}
\caption{Normalized redshift distributions for the 55{,}826 \textsc{ngc-core}, 30,551 \textsc{ngc-bonus} 
and 86,377 \textsc{ngc-core+bonus} quasars over the redshift range $2.2 \le z \le 2.8$.  Also indicated is the redshift distribution generated for
the random catalog that is used to mimic  \textsc{ngc-core} quasars.}
\label{zhistnorm}
\end{center}

\end{figure}

\subsection{Targeting Completeness}

Quasar clustering measurements are sensitive to the angular completeness of the
quasar sample (the ``mask''). 
Quasar spectra in \textsc{boss}
were obtained using a set of spectroscopic tiles that can be modeled as  
polygons using the {\sc mangle} software \citep{bl03,teg04,sw08}.
The fraction of targets (quasars) in an
area of the sky that has been covered by a unique set of these tiles (called a
sector) is used to define the completeness of the sample. Following
\citet{wh12}, we limit the survey to sectors with targeting completeness ($f_{\rm comp}$) greater
than 75\%. About 60\% of such sectors are 100\% complete, and the mean
area-weighted completeness of the sectors we use is $\sim95$\%.
A veto mask is applied to remove the regions in which quasars 
can not be observed such as around bright stars, and near the centerposts of the
spectroscopic plates \citep{wh12}.

Note that fiber collisions make it impossible to obtain spectra for quasar
pairs closer than 62$\arcsec$ except in overlapping regions \citep{bl03,wh12}.
At the mean redshift of our survey ($\bar{z}\simeq 2.43$), $62\arcsec$
corresponds to a transverse comoving distance of $1.2\,h^{-1}$Mpc.
We have not corrected for this fiber collision effect, as it is significant
only near the lower edge of the range of scales we consider; this is a region where the errors from shot noise are
already quite large.

We adopt the spectroscopic redshift reported by the spectral reduction pipeline
 {\tt spAll-v5\_7\_0} to compute the comoving distances. 
To avoid using any target with a problematic redshift, we only accept quasars with
z\textsc{warning}=0, indicating a confident spectroscopic 
classification and redshift measurement for quasar targets \citep{bol12}\footnote{http://www.sdss3.org/dr9/spectro/caveats.php\#zstatus}. 

Unless
otherwise specified in the text, we also limit our clustering measurements to just
the {\sc ngc} region of Figure \ref{aitoff}, as the {\sc ngc} represents
the largest contiguous area in {\sc boss}.
Excluding the {\sc sgc} quasars from the original overall \textsc{core} sample
of 69{,}977 quasars significantly affected the clustering signal both in real
and redshift space.
Previous analyses have also found unexplained differences between clustering
measurements
in the \textsc{boss} \textsc{ngc} and \textsc{sgc} regions \citep[e.g.,][]{wh12}.
The target selection papers for the coming e\textsc{boss} survey
(Myers et al. 2015, Prakash et al. 2015, in preparation)
suggest that the \textsc{sgc} contains areas in which true target density is difficult to regress,
which may explain such discrepancies.
Table \ref{tabstat} documents how our available sample sizes are affected
by restricting to the \textsc{ngc}, to completenesses of $f_{\rm comp}>0.75$,
and to z\textsc{warning}=0.

\begin{figure}
\begin{center}
 \resizebox{\columnwidth}{!}{\includegraphics{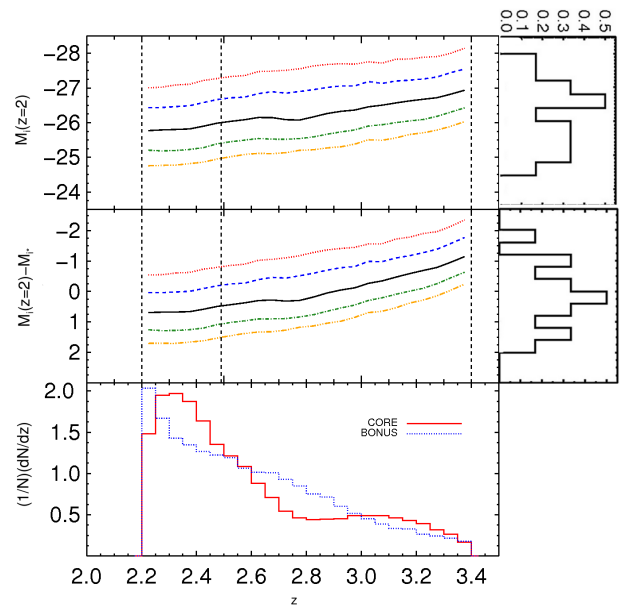}}
\caption{The absolute magnitude distribution for \textsc{boss} quasars. 
The upper panel displays the $10^{\rm th}$, $25^{\rm th}$, $50^{\rm th}$, 
$75^{\rm th}$ and $90^{\rm th}$ percentiles of $M_{i}$ vs.\ redshift. The right-hand side histogram for the upper panel is the normalized distribution of $M_{i}$ for the full \textsc{ngc-core} sample over $2.2 \le z \le 3.4$.
The middle panel shows the same percentiles for $M_{i}- M_{*,i}$ vs. redshift, where
$M_{*,i}$ is the characteristic luminosity derived from Eqn.\ \ref{mi}. Similar to the upper panel, the right-hand side histogram for the middle panel is the normalized distribution of $M_{i}- M_{*,i}$ for the full \textsc{ngc-core} sample over $2.2 \le z \le 3.4$. The three vertical dashed lines 
depict the lower limit, median and the upper limit for the full redshfit range of the samples studied in this paper (2.2, 2.49 and 3.4 respectively).
The lower panel is the normalized redshift distribution of quasars over the range $2.2 \le z \le 3.4$.}
\label{absmag}
\end{center}
\end{figure}

\begin{table}
\begin{tabular}{ll}

\hline
\hline
Properties & \# of quasars \\
\hline
\underline{XDQSO targets:} & \\
In the \textsc{core} sample & 107455 \\
In the \textsc{bonus} sample & 104971 \\
& \\
\underline{Targets with z\textsc{warning}=0:} & \\
In the \textsc{core} sample &  103552 \\
In the \textsc{bonus} sample & 89044 \\

& \\
\underline{Targets with $f_{\rm comp}> 0.75$ and z\textsc{warning}=0:} & \\
In the \textsc{core} sample &  97644 \\
In the \textsc{bonus} sample & 79699 \\
\textsc{ngc}\footnote{North Galactic Cap } \textsc{core} sample & 78151\\
\textsc{ngc} \textsc{bonus} sample & 60788\\
& \\
\underline{$f_{\rm comp}> 0.75$, $2.2 \le z \le 2.8$:} & \\
In the \textsc{core} sample &  70892 \\
In the \textsc{bonus} sample &  44786\\

& \\
\underline{$f_{\rm comp}> 0.75$, $2.2 \le z \le 2.8$, z\textsc{warning}=0:} & \\
In the \textsc{core} sample &  69977 \\
In the \textsc{bonus} sample &  41019 \\
\textsc{ngc} \textsc{core} sample & 55826 \\
\textsc{ngc} \textsc{bonus} sample & 30551 \\
& \\
\underline{$f_{\rm comp}> 0.75$, $2.2 \le z \le 3.4$:} & \\
In the \textsc{core} sample &  94306 \\
In the \textsc{bonus} sample & 60277 \\
& \\

\underline{$f_{\rm comp}> 0.75$, $2.2 \le z \le 3.4$, z\textsc{warning}=0:} & \\
In the \textsc{core} sample &  92347 \\
In the \textsc{bonus} sample & 54614 \\
\textsc{ngc} \textsc{core} sample & 73884\\
\textsc{ngc} \textsc{bonus} sample & 40781 \\
\hline
\end{tabular}
\caption{Number of objects in the SDSS DR12 BOSS XDQSO sample with different cuts on redshift and completeness.}
\label{tabstat}
\end{table}

\subsection{Division by Redshift and Luminosity}
\label{sec:lumdata}

To investigate the evolution of quasar clustering, we extended the redshift 
range by relaxing the upper limit from 2.8 to 3.4 and split the 
\textsc{ngc-core} sample into three redshift bins that are populated by nearly 
the same number of quasars (see Table \ref{tabz}). We also check whether the 
clustering signal is
luminosity dependent by dividing the \textsc{ngc-core} sample into three
luminosity subsamples based on the absolute magnitude of the quasars in the
$i$-band. Following \citet{ro13}, we $k$-correct 
all of our quasar magnitudes to $z=2$.
Figure \ref{absmag} presents the absolute magnitude distribution of the
quasars in the final {\sc boss} \textsc{ngc-core} sample with respect to the
characteristic luminosity of quasars at every redshift in this range:
\begin {equation}\label{mi}
 M_{i,*}(z)=-21.61-2.5~ (k_{\rm 1}z+ k_{\rm 2}z^{2})-0.71 ~ , ~ 
\end{equation} 
where $k_{\rm 1}=1.39$ and $k_{\rm 2}=-0.29$ for $z \le 3$ and $k_{\rm 1}=1.22$
and $k_{\rm 2}=-0.23$ for $z \ge 3$ (from \citealt{cr04}, modified by
\citealt{crot06}). Quasars at higher redshift
are fainter, and fainter quasars have larger flux uncertainties which translate
into imprecision in measured colours. So, at higher redshift the $k$-correction
becomes increasingly uncertain. 
\citep[e.g.][]{rich06}.

\section{Clustering Measurements and Error Analysis}\label{cls}

To determine how much the distribution of quasars deviates from a
homogeneous distribution, it is compared to a ``random'', or control,
set of objects. This random catalog has all of the characteristics of
the data, including the selection function, except for its clustering.
Figure \ref{zhistnorm} shows how the redshift distribution of the random
catalog we have produced mimics that of the final \textsc{core} {\sc boss} quasar sample.
Appendix A introduces the code that we use to make the random catalog from
the survey mask and quasar redshift distribution.

The two-point correlation function $\xi(x)$ is defined as the probability,
in excess of random, of finding a pair of objects at a separation $x$:
$\delta P=n\delta V[1+\xi(x)]$ \citep{pe80}.
We use the estimator introduced by \citet{ls93} to calculate the real
and redshift-space two-point correlation functions (2PCFs), $\xi(x)$,
of quasars in our sample:
\begin{equation}\label{xifrac}
 \xi(x)= \frac{\langle DD(x)\rangle-2\langle DR(x)\rangle + \langle RR(x)\rangle}{\langle RR(x)\rangle},
\end{equation}
where $DD(x)$ is the number of data-data pairs with separation between $x$ and
$x$+$\Delta x$ and $DR(x)$ is the number of data-random pairs also separated by
$x$. The random catalog is constructed to be larger than the data catalog to reduce
Poisson noise due to the pair counts that include random points. Consequently, there are normalization factors of the form $N_{\rm
R}/N_{\rm D}$ that scale the counts appropriately if the random set has a
different size than the data set. The angled brackets denote the {\em
suitably averaged} pair counts.  Throughout this paper we use random catalogs 12 
to 20 times larger than our quasar samples. Tests with random catalogs
up to 30 times larger than the data show little statistical improvement in the
clustering signal, but increase the computing time for our analyses. 

Using the comoving distances to each of the two objects in a pair ($s_{1}$ and
$s_{2}$), and the angular separation between them ($\theta$), we define each pair's
comoving separation along ($\pi=|s_{1}-s_{2}|$) and across 
($r_{\rm p}=[s_{\rm 1}+s_{\rm 2}] \theta/2$) the line of
sight\footnote{Also denoted by $\sigma$ or $\rm R$ in the literature.}.
We thus derive the two-dimensional correlation function $ \xi(r_{\rm p},\pi)$.
Since the effects of redshift space distortions are limited to the line
of sight (radial) direction, we can integrate $ \xi(r_{\rm p},\pi)$ over $\pi$
in order to eliminate the redshift-space distortion effect:

\begin{equation}
\label{wpint}
w_{\rm p}(r_{\rm p})=2 \int^{\pi_{\rm cut}}_{0}  \xi(\pi,r_{\rm p}) \rm d\pi ,
\end{equation}

where the upper limit, ${\pi_{\rm cut}}$, is the distance at which the
effects of redshift-space distortions become negligible\footnote{The quantity 
$w_{\rm p}(r_{\rm p})$ integrates out the effects of {\em both}
peculiar velocities and  redshift errors \citep[e.g.,][]{cro05,da05,wh12}.}.
We refer to $w_{\rm p}$ as the projected correlation function.
Averaging over all pairs whose total redshift-space separation lies
in a bin produces the monopole moment of the redshift-space correlation
function, $\xi(s)$.

The size, bias and number density of the sample together with the redshift
range over which the clustering signal is being measured, can change the optimum
value of $\pi_{\rm cut}$.
For instance, $70\,h^{-1}$Mpc was used for the 2dF QSO survey of
23{,}338 quasars \citep{da05}, $\sim 100\,h^{-1}$Mpc was used by
\citet{she07} for 6{,}109 quasars at $z \ge 2.9$, and
$50\,h^{-1}$Mpc was used for the sample of 27{,}129 DR10 quasars
in the \textsc{sdss-iii/boss} survey \citep{wh12}.
 The value of $\pi_{\rm cut}$ should be chosen to balance the advantage of integrating out
redshift-space distortions against the disadvantage of introducing noise from
uncorrelated line-of-sight structure.
Moreover, integrating over a wider range of $\pi$ means projecting a larger
3D space into smaller $r_{\rm p}$ bins \citep[e.g.,][]{wh12}.
 After testing a wide enough range of upper 
limits to truncate the integral, we found $\pi_{\rm cut}=50\,h^{-1}\rm {Mpc}$ to 
be the optimum limit beyond which the 
$ \xi(r_{\rm p},\pi)$ values become negligible for our sample.

We estimate the uncertainties using inverse-variance-weighted jackknife
resampling \citep{sc02,ze02,my05}.
We divide our sample into $N$ pixels on the sky, then create $N$ subsamples
by neglecting pixels one-by-one and using the remaining area to calculate
the 2PCF.
We pixelize our sample into angular regions that
are specified by HEALPIX\footnote{http://healpix.sourceforge.net/} pixels
\citep{gr05} with $N_{\rm side}=4$ ($\sim 15^{\circ}$ on a side).
To compensate for missing areas, we merge pixels with fewer random points
than two-thirds of the mean\footnote{ Although this process could, in principle, be performed multiple times, we only merged the pixels once.}
The inverse-variance-weighted covariance matrix elements, $C_{ij}$, are

\begin{multline}\label{covmatrix}
C_{ij}=C(r_{i},r_{j})=\sum^{N}_{L=1}\frac{RR_{L}( r_{i})}{RR(r_{i})} [\omega_{L}(r_{i})-\omega( r_{i})] \\
\times \frac{RR_{L}(r_{ j})}{RR(r_{j})} [\omega_{L}(r_{j})-\omega(r_{j})],
\end{multline}
where the $L$ subscript refers to the neglected pixel and the random pair
ratio, $RR_{L}/RR$, compensates for the relative area through the relative
random catalog size in each pixel \citep[e.g][]{my07}.
One-standard-deviation jackknife errors, $\sigma_{i}$, are the square root
of the diagonal elements of the covariance matrix.
We assume the errors are Gaussian, and thus compute the goodness-of-fit or
likelihood of any model that fitted to the data through
\begin{equation}\label{chi2}
  \chi^{2}=\sum_{i,j}\left[w(r_{i})-w_{\rm model}(r_{i})\right]
  C^{-1}_{ij} \left[w(r_{j})-w_{\rm model}( r_{j})\right]  .
\end{equation}

\section{Clustering Results} \label{clsres}

\subsection{Real-space correlation function}\label{realACF}

To measure the projected auto-correlation function $w_{\rm p}(r_{\rm p})$
we use a sample of 55{,}826 \textsc{ngc-core} quasars within the redshift
range $2.2 \le z \le 2.8$ 
(see Table \ref{tabstat}).

For sufficiently large $\pi_{\rm max}$, $w_{\rm p}(r_{\rm p})$ can be
related to the real-space clustering as \citep{dp83}
\begin{equation}\label{re}
w_{\rm p}(r_{\rm p})=2 \int^{\infty}_{r_{\rm p}}
  \frac{r \xi(r)}{\sqrt{ r^{2}-r^{2}_{\rm p}}} {\rm d}r.
\end{equation}
As a first model, we use a simple power law of the form
($r/r_{\rm 0})^{-\gamma}$ for the real-space correlation function. 
For this power-law form, Eqn.\ \ref{re} reduces to 
\begin{equation}\label{frw}
  \frac{w_{\rm p}(r_{\rm p})}{r_{\rm p}}=\frac{\sqrt{\pi}~
  \Gamma[(\gamma-1)/2]}{\Gamma[\gamma/2]}
  \left(\frac{r_{\rm 0}}{r_{\rm p}}\right)^{\gamma} .
\end{equation} 
We approach the fitting process in two separate ways: 
\par
(1) First, we perform a two-parameter fit using the $\chi^{2}$ minimization method to determine
both the slope and the intercept of the fitted line. We use the resulting
slope ($\gamma \sim 1.85$) as a fixed parameter and perform a one-parameter
$\chi^{2}$ minimization, obtaining 
a scale length of $\rm r_{0}=7.7 \pm 0.2\,h^{-1}{\rm Mpc}$ with 
$\chi^{2}_{\rm red}=0.36 $ over 9 degrees of freedom.

(2) Second, we fix $\gamma=2$ and fit a 1-parameter model. This allows us
to evaluate how much clustering amplitude is sensitive to the chosen slope of the model.
This fit results in a scale length of  
$r_{\rm 0}=8.12\pm0.22\,h^{-1} {\rm Mpc}$ with $\chi^{2}_{\rm red}=0.25 $ over 10 degrees of freedom.

Since a slight change in the slope does not make a significant difference to the result of the fit for the clustering amplitude, 
we conclude that the clustering amplitude for our quasar sample in the fitted range is not 
strongly sensitive to the slope of the real-space correlation function. As we find that $\gamma=2$ is an acceptable value of the slope, we proceed by using $\gamma=2$ throughout the rest of this paper so that our fitted correlation lengths can be compared more easily to other works \citep[e.g][] {she07,wh12}.    

\begin{figure}
\begin{center}
\includegraphics[angle=0,scale=0.31]{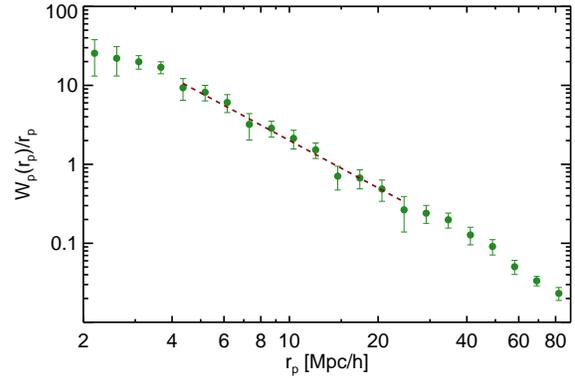}
\caption{The projected auto-correlation function for the \textsc{core} sample
over the redshift range $2.2 \le z \le 2.8$.
 The dashed line is the best-fit one-parameter power law over the range $4
< r_{\rm p} < 25\,h^{-1} {\rm Mpc}$. Because the two-parameter
power-law fit does not have a
 significantly different slope to the one-parameter power-law over the chosen
fitting range, we adopt $\gamma=2$ throughout this paper for consistency with
\citet{wh12}.}

\label{ACFfit}
\end{center}
\end{figure}

Fig.\ \ref{ACFfit} presents the results of the fits to the measured
$w_{\rm p}$.
The fitting range in both approaches is $4 \lesssim r_{\rm p} \lesssim
25\,h^{-1} {\rm Mpc}$. We find that  $\chi^{2}$ maintains an acceptable
level, and that our fitting results are consistent within the derived uncertainties, for all fitting ranges from  $4 \lesssim r_{\rm p}
\lesssim 16\,h^{-1} {\rm Mpc}$ to $2 \lesssim 
 r_{\rm p} \lesssim 50\,h^{-1} {\rm Mpc}$.
An improved mask and/or deeper imaging for the initial target selection of rare and
faint high-redshift quasars would be needed to push the fitting scales further in
order to measure subtle large-scale features, but the quasar bias measurement
should be well characterized 
over the moderate scales of $< 25\,h^{-1} \rm Mpc$.

\begin{table*}
\begin{center} 
\begin{tabular}{lccccccccccc}
\hline
\hline
$r_{\rm p}$   &    4.36    &   5.18   &    6.15   &    7.31   &    8.69   &    10.33    &   12.28   &    14.59   &    17.34   &   20.61  &     24.50 \\

\hline 
\\
$w_{\rm p}$ &  54.64   &    48.87   &    43.41    &   38.09    &   29.77   &    26.47     &  22.74    &   16.68    &   11.90    &   15.27   &    7.80  \\

\hline
\\
$\sigma$      &    12.45   &    9.45   &    9.56    &   8.64    &   5.68    &    5.88    &   4.17  &     3.44   &   3.16   &    3.01   &    3.09  \\

\hline

\\
4.36    &    1.000  &   -0.256  &  -0.052  &   -0.343   &   0.027    &   0.485  &   0.221   &   0.171   &    0.174   &    0.553    &   0.543  \\
5.18     &       -       &    1.000  &   0.023   &   0.152  &   0.455  &   -0.215  &   0.320  &   -0.298   &   -0.299   &   -0.245    &  -0.241  \\
6.15     &       -       &          -    &    1.000   &    0.232  &    -0.119   &    -0.379   &   -0.083    &   -0.225     &   -0.261     &   -0.419    &  -0.273  \\
7.31     &       -       &        -      &        -    &     1.000   &   0.083  &    -0.410   &   0.095   &   -0.050    &   -0.958    &   -0.606     &  -0.579  \\
8.69     &      -        &        -      &        -        &       -       &       1.000  &    0.472   &   0.202  &    0.153   &    0.656   &    0.538    &   0.529  \\
10.33     &      -        &         -      &        -        &       -        &       -      &      1.000   &    0.484   &    0.382    &    0.388     &   0.138      &  0.119  \\
12.28     &      -        &         -      &        -        &       -        &       -        &       -      &      1.000   &   0.031   &    0.035     &  -0.166     &  0.154  \\
14.59     &      -        &         -      &        -        &       -        &       -        &       -      &        -      &      1.000    &    0.004   &    0.323    &   0.309  \\
17.34    &      -        &         -      &        -        &       -        &       -        &       -      &       -        &       -       &     1.000    &   0.320     &  0.306  \\
20.61     &      -        &         -      &        -        &       -        &       -        &       -      &       -        &       -        &       -      &     1.000     &  0.083  \\
24.50     &      -        &         -      &        -        &       -        &       -        &       -      &       -        &       -        &       -       &     -      &        1.000  \\

\hline
\end{tabular}
\end{center}
\caption{Correlation coefficients as estimated from the covariance matrix 
computed by jackknife error estimation for \textsc{ngc-core} quasars. The first 
three rows are the comoving separation, $r_{\rm p}$ (in
$h^{-1} {\rm Mpc}$), the auto-correlation $w_{\rm p}$ and its uncertainty.
The remainder of the table lists the correlation coefficients as estimated from
the covariance matrix computed by jackknife error estimation.}
\label{tabcov}

\end{table*}

Using the method described in $\S$\ref{cls}, we estimate the covariance matrix of the 
auto-correlation function (ACF) for \textsc{ngc-core} quasars. The $w_{\rm p}$ result for 
the ACF of the \textsc{ngc-core} quasar sample over the fitting range 
$4 < r_{\rm p} < 25\,h^{-1} {\rm Mpc}$ and its uncertainty ($\sigma$) 
are reported in the first three rows of Table \ref{tabcov}. The remainder of the table lists the 
correlation coefficients as estimated from the covariance matrix calculated 
via inverse-variance-weighted jackknife resampling.

\subsection{Using \textsc{boss} \textsc{bonus} quasars}
\label{sec:bonus}

As mentioned in $\S$\ref{dat}, \textsc{boss} also surveyed a heterogeneously selected
quasar sample with a complicated angular selection function---the \textsc{bonus} sample.
Although modeling the mask for the \textsc{bonus} sample is beyond the scope of this
paper, we can cross-correlate a \textsc{bonus+core} sample against the \textsc{core} 
sample, using the random catalog constructed for the \textsc{core} sample, in an attempt 
to improve our measurement precision. Expanding the main sample by
adding quasars from the \textsc{bonus} sample might be expected to improve the clustering signal (as
there are 
more total pair counts in a \textsc{core+bonus}-to-\textsc{core} cross
correlation than in a \textsc{core}-to-\textsc{core} ACF) but it is worth
determining whether such an approach is
justified by improved statistics.
 
\begin{figure}
\begin{center}
\includegraphics[angle=0,scale=0.34]{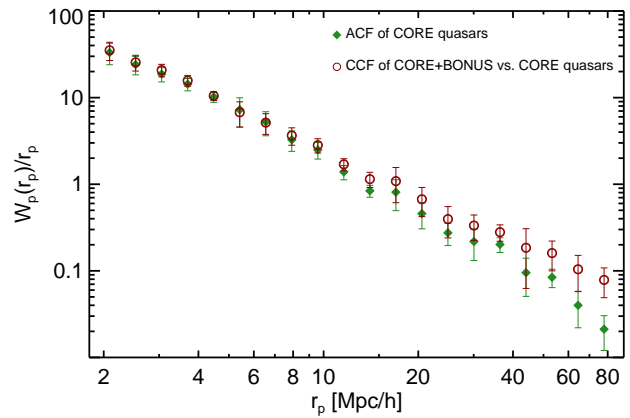}
\caption{Projected cross-correlation between the 55,826 \textsc{core} and 86,377
\textsc{core+bonus} quasars over the redshift range $2.2 \le z \le 2.8$ (red
open circles). The cross-correlation result is consistent with the
auto-correlation of the \textsc{core} sample (green filled circles) over our
scales of interest.
As little additional precision is introduced by using the full \textsc{core+bonus}
sample, we restrict our
analysis to just the auto-correlation of the \textsc{core} sample.}
\label{ACF-CCF}
\end{center}
\end{figure}

Fig. \ref{ACF-CCF} shows the cross-correlation of \textsc{core+bonus} (86{,}377
quasars) against \textsc{core} (55{,}826 quasars) for objects in the \textsc{ngc}
with z\textsc{warning}=0 and completeness greater than 75$\%$. 
 For the cross-correlation estimator, the ``DD'' pairs in Eqn.\ 
\ref{xifrac} consist of one member from the \textsc{core} and another from the
\textsc{core+bonus} sample. The random catalogue, here, was made for the uniformly selected (i.e. \textsc{core}) sample, so the ``2DR'' pair breaks into a ``DR'' pair for the \textsc{core} and a ``DR'' pair from the \textsc{core+bonus} sample, where "R" always denotes the random catalog constructed for the CORE sample. In other words, Eqn. \ref{xifrac} for the cross-correlation estimator that we adopt is $\xi(x)=(\langle D_{C} D_{CB}(x)\rangle-\langle D_C R(x)\rangle -D_{CB} R(x)+ \langle RR(x)\rangle)/\langle RR(x)\rangle $, where the subscripts \textsc{``c''} and \textsc{``cb''} refer to the \textsc{core} and \textsc{core+bonus} samples respectively. The appropriate data/random normalization factors are included here (as in the description of Eqn.\ \ref{xifrac}).

The difference
between the ACF and cross-correlation 
at every scale is well within the $1 \sigma$ uncertainty. However, the cross
correlation has 
slightly better precision on smaller scales, which would be a benefit to
measurements of, e.g., the one-halo term in the HOD formalism. 
The consistency of the clustering replacing the ACF with a
cross correlation is a useful check on our analyses, and may also represent one way forward for performing clustering measurements using heterogeneous samples. 
However, given the only minor improvement, we continue using the ACF throughout this paper. 

\begin{table*}
\begin{center} 
\begin{tabular}{cccccccc}
\hline
\hline

$r_{\rm p}$ &  $w_{\rm p}$ & $\sigma$& $w_{\rm p}$ & $\sigma$ & $s$ & $\xi(s)$ &  $\sigma$ \\ 
$(h^{-1} {\rm Mpc})$ & ACF  &  & CCF &  & $(h^{-1} {\rm Mpc})$ &  &  \\ \hline

 2.09  &        69.353 &        19.294 &        73.474 &        17.498  &      3.06  &      2.556  &     0.609   \\
 2.53 &        60.807 &        14.491 &        64.523 &        13.311     &      3.70  &      1.663  &     0.384  \\
 3.06 &        57.125 &        10.707 &        63.607 &        10.438 &      4.48  &      2.406  &     0.318  \\
 3.70 &        52.908 &        8.654 &        58.216 &        8.133 &      5.41  &      1.654  &     0.213  \\
 4.47 &        45.454 &        5.981 &        46.928 &        5.365 &      6.55  &      1.020  &     0.132  \\
 5.41 &        39.291 &        10.607 &        36.604 &        11.831      &      7.92  &      1.034  &     0.101  \\
 6.55 &        34.392 &        9.142 &        33.677 &        9.015 &      9.59  &     0.676  &      0.069 \\
 7.92 &        25.846 &        6.051 &        28.909 &        6.567 &      11.60  &     0.562  &      0.050  \\
 9.58 &        23.907 &        5.210 &        27.066 &        5.061 &      14.03  &     0.433  &      0.036 \\
 11.60 &        16.056 &        3.002 &        19.692 &        3.129 &      16.97  &     0.291  &      0.026  \\
 14.03 &        11.771 &        1.854 &        16.066 &        3.196 &      20.54  &     0.193  &      0.018 \\
 16.97 &        13.753 &        5.348 &        18.433 &        8.045 &      24.84  &     0.150  &      0.014  \\
 20.54 &        9.404 &        3.130 &        13.768 &        5.087 &      30.06  &    0.090  &       0.010  \\
 24.85 &        6.824 &        0.958 &        9.8349 &        3.872  &      36.37  &    0.069  &      0.007  \\
 30.06 &        6.554 &        2.099 &        10.018 &        3.256 &      43.10  &    0.040  &       0.006  \\
 36.37 &        7.341 &        1.422 &        10.170 &        2.184 &      53.23  &    0.020  &       0.004  \\
 44.00 &        4.192 &        1.061 &        8.122 &        5.365 &      64.40  &    0.010  &       0.003  \\
 53.23 &        4.483 &        1.076 &        8.535 &        3.208 &      77.92  & -0.000  &     0.002  \\ 
 64.40 &        2.576 &        0.159 &        6.711 &        3.004  &      94.27  &   0.004  &        0.002  \\
\hline
\end{tabular}
\end{center}

\caption{Projected and redshift-space correlation functions. ACF $w_{\rm p}$ represents the auto-correlation of the
\textsc{ngc-core} sample, CCF $w_{\rm p}$ is the cross-correlation between the
\textsc{ngc-core+ngc-bonus} and \textsc{ngc-core} samples and $\xi(\rm s)$ is
the redshift-space correlation for \textsc{ngc-core} quasars over the redshift
range $2.2 \le z \le 2.8$.}

\label{tabwpxi}
\end{table*}

\subsection{Redshift-space 2-point correlation function}\label{zsp}

Using redshift to infer distance for the 2PCF causes a complication; peculiar 
velocities introduce redshift-space distortions in $\xi$ along the line of sight \citep{kai87}.
This effect has utility, though, as
measuring these redshift space distortions can constrain cosmological parameters.

\begin{figure*}
\begin{center}
\includegraphics[angle=0,scale=0.55]{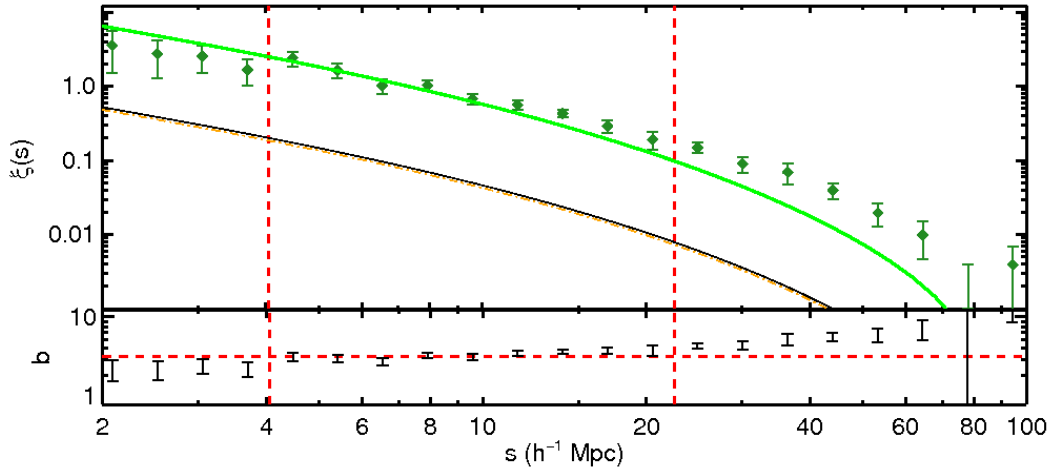}
\caption{The redshift-space correlation function $\xi(s)$ for 55{,}826 {\sc
boss} \textsc{ngc-core} quasars over the redshift range $2.2 \le z \le 2.8$. The
black curve is the real-space correlation function, $\xi(r)$ of dark matter from
{\sc halofit} \citep{sm03} modified by the correction
for redshift space distortions from Kaiser (1987). The resulting best-fit bias
(green curve) is $b_{\rm Q}$ = 3.54. The lower panel depicts the bias residuals compared to
the fiducial fit. The (red) vertical dashed lines at $\sim 4\,h^{-1} \rm Mpc$ and $\sim 22\,h^{-1} \rm Mpc$
depict the scale over which we fit the data.}
\label{xi}
\end{center}
\end{figure*}

\begin{figure}
\begin{center}
\includegraphics[angle=0,scale=0.26]{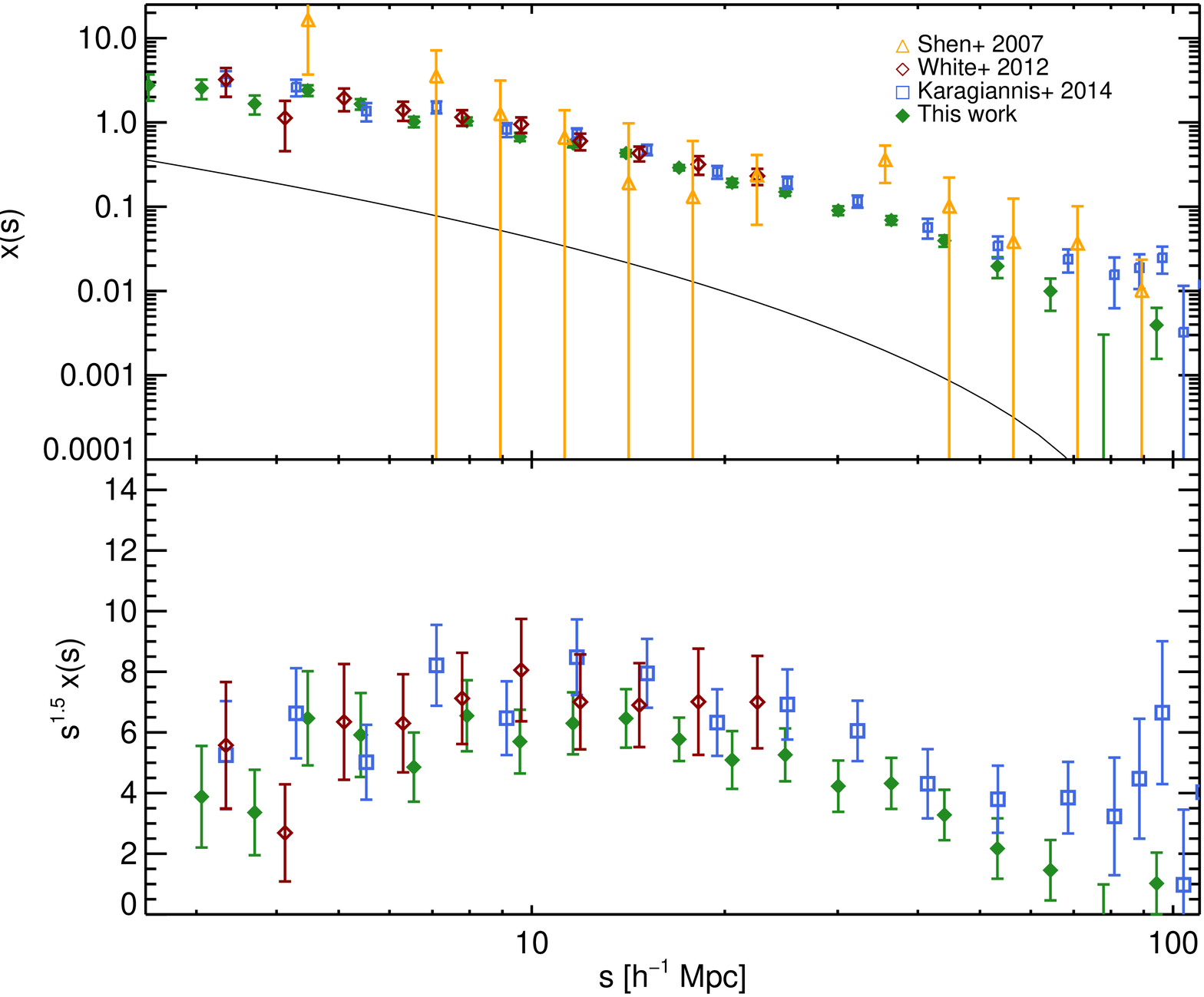}
\caption{Comparison of our redshift-space correlation function with earlier
works. \citet{wh12} and \citet{kar14} both used
samples of $\sim25{,}000$ \textsc{core} quasars from \textsc{boss} over the same
redshift range we
consider in this paper ($2.2 \le z \le 2.8$). \citet{she07} used $\sim$4{,}400
quasars from \textsc{sdss} Data Release 5 over $2.9 \le z < 5.4$. The black
curve is as for
Fig.\ \ref{xi}.}
\label{xicomp}
\end{center}
\end{figure}

Fig.\ \ref{xi} shows the redshift-space correlation function $\xi(s)$ of 55{,}826
\textsc{ngc-core} quasars over the redshift range 2.2 $\le z \le $ 2.8. As
described in $\S$\ref{cls}, a jackknife resampling is used to derive both the
uncertainties and the full covariance matrix. 
The black curve is the real-space correlation function, $\xi(r)$ of dark matter
computed from {\sc halofit} \citep{sm03} modified by the 
relationship between the redshift-space and real-space correlation functions
\citep{kai87}
\begin{equation}
\label{xis_xir}
\xi(s)=\left(b^{2}+\frac{2}{3}bf+\frac{f^{2}}{5} \right) \xi(r),
\end{equation} 
where $f=[\Omega_{m}(z)]^{0.56}$ is the gravitational growth factor. The model 
is then fit to the measured correlation function for quasars by $\chi^{2}$ minimization. 
The best-fit bias of quasars relative to the underlying dark matter is 
$b_{Q}$ = 3.54 $\pm$ 0.1 with  $\chi^2_{\rm red}=1.06$ for 7 degrees of freedom 
using the full covariance matrix. The lower panel of Fig. \ref{xi} presents the
bias residuals and the horizontal dashed line represents the best-fit value of
the bias. The vertical dashed line 
indicates the edge of the last bin that is included in the fit. This figure
demonstrates that 
scale-independent bias combined with a $\Lambda$CDM cosmology is a sufficient 
fit to our data over scales of at least $4 \lesssim s \lesssim 22\,h^{-1} {\rm Mpc}$. 

Table \ref{tabwpxi} presents our measurements of the projected auto-correlation
function and redshift-space auto-correlation function, as well as the projected
cross correlation with \textsc{bonus} quasars (see \S\ref{sec:bonus}). 
The reported uncertainties for all correlation functions are the diagonal
elements
of the covariance matrix derived by inverse-variance-weighted jackknife
resampling.
Our result for the redshift-space correlation function is compared to other works in Fig.\ \ref{xicomp}.
The figure shows some consistency between previous results, although our work is in better
agreement with a $\Lambda$CDM cosmology on large scales.

 In general, the excess of the data above the model seen by ourselves and
other authors for scales greater than $\sim 25 \,h^{-1} {\rm Mpc}$ is likely to
be due
to the difficulty in building a random catalog that fully mimics complex and
subtle deviations in target density that manifest on large scales where the
correlation amplitude is small \citep[e.g.,][]{ross11, ho15}.
Notably, \citet{ag14} demonstrate that {\em unknown} systematics can
affect quasar clustering amplitude on very large scales.
The disagreement between our adopted model and the data on scales of $s >
27 \,h^{-1} {\rm Mpc}$ is likely to be due to the
very small error bars on these scales being driven by correlations in imaging
systematics, which, when jackknifed, lead to an
underestimate of the true errors. Weighting the target density using systematics
maps to see if the
clustering fit can be pushed to larger scales is currently under investigation,
but is beyond the scope of this
paper. CDM is known to be a good description of underlying dark matter and,
therefore, in this work, we
focus on deriving the quasar clustering amplitude and slope for the scales on
which CDM represents a good fit, and on which we therefore
believe our data to be uncontaminated by subtle systematics.

\subsection{2-D Redshift-space correlation function, $\xi(r_{\rm p},\pi)$}\label{zsp2d}

\begin{figure}
\resizebox{\columnwidth}{!}{\includegraphics{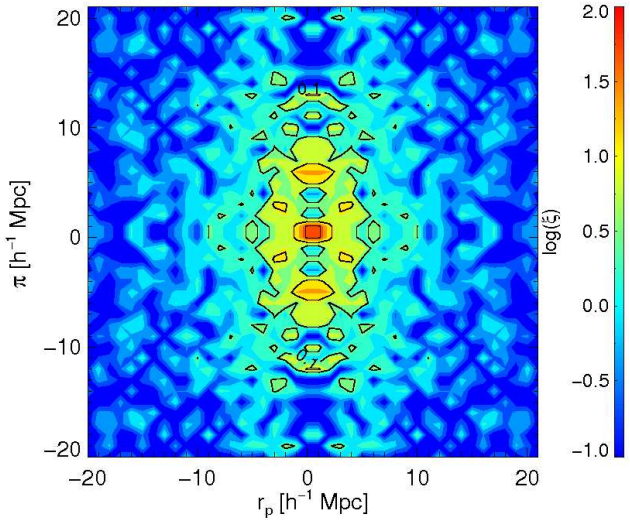}}
\caption{
The two-dimensional redshift-space correlation function
$\xi(r_{\rm p},\pi)$ for 55{,}826 \textsc{ngc-core boss} quasars.
The extension in the line-of-sight direction is due to a combination of
redshift errors and ``Fingers-of-God''.}
\label{contour}
\end{figure}

As mentioned in $\S$\ref{cls}, one can project the redshift space separation
between two objects into orthogonal components along ($\pi$) and across ($r_{\rm
p}$) the line of sight. 
The elongation 
of virialized overdensities such as clusters of galaxies along the line of
sight, caused by small peculiar velocities that are not associated with the
Hubble flow, should be 
visible in 2-D redshift-space maps as long, narrow filaments aligned with the line of sight \citep{teg04}. To check whether this
``Fingers of God'' effect is apparent in our sample, 
we measure $\xi(r_{\rm p},\pi)$ for our \textsc{ngc-core} quasars out to scales 
of $\pm 20\,h^{-1} \rm Mpc$, and display the results 
in Fig.\ \ref{contour}. It is clear from Fig.\ \ref{contour} that the redshift-space distortions are strongest for
$\pi < 20\,h^{-1}\rm {Mpc}$, confirming that our adopted redshift-space integration limit of 
$\pi_{\rm cut}=\pm 50\,h^{-1}\rm {Mpc}$ is reasonable. Although we choose not to further analyze 
$\xi(r_{\rm p},\pi)$ in this paper, Fig.\ \ref{contour} implies that studies of
redshift-space distortions would be a meaningful avenue for further cosmological studies
using \textsc{boss} quasars.



\section{Evolution and luminosity dependence of quasar clustering}\label{ev}

\begin{figure*}
\begin{center}
\includegraphics[angle=0,scale=0.285]{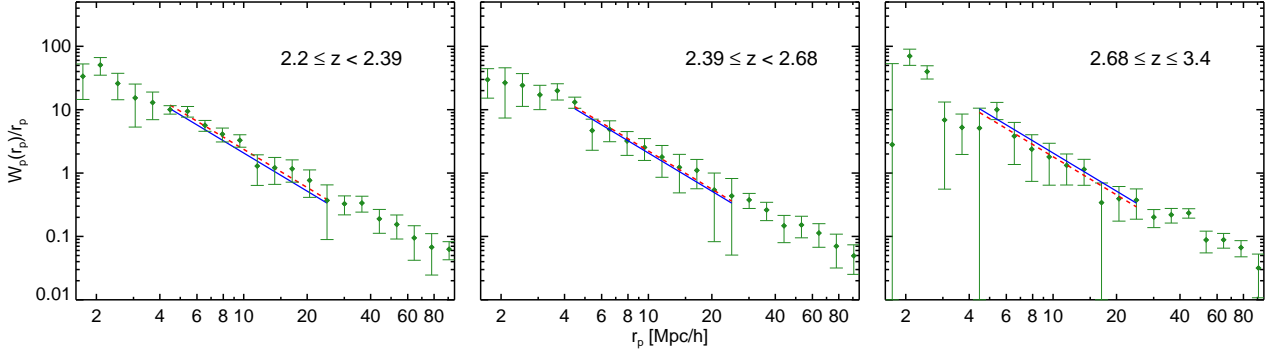}
\caption{The projected real-space clustering of the \textsc{ngc-core} sample 
in three bins of redshift containing roughly equal numbers of quasars.
See Table \ref{tabz} for the sample properties and Table \ref{tabrb} for
the fitting results. The red dashed line is the one-parameter fitted power-law with $\gamma=2.0$ and the blue solid line is the one-parameter fitted power-law to the measured correlation function of the main sample plotted here as a reference line. }
\label{wp_3subz}
\end{center}
\end{figure*}

\begin{figure*}
\begin{center}
\includegraphics[angle=0,scale=0.4]{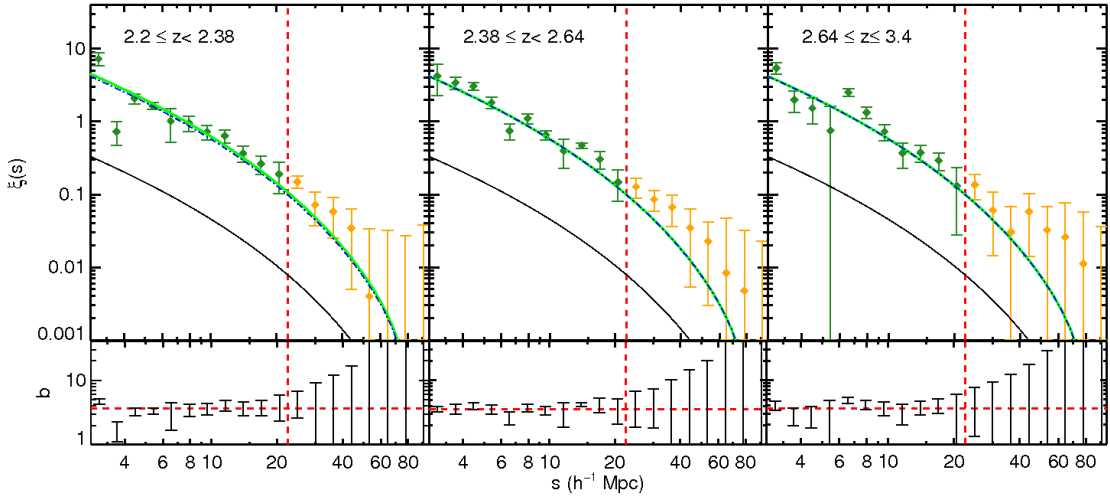}
\caption{The measured redshift-space correlation function $\xi(s)$ for the
three redshift subsamples detailed in Table \ref{tabz}. 
The fitting range for all three subsamples is
$3.0 \lesssim s \lesssim 22.5\,h^{-1} \rm Mpc$ and the fitting results are 
detailed in Table \ref{tabrb}. The green curve is the fitted dark matter
correlation function to the measured $\xi(s)$ for each sample and the overlaid blue dashed curve
is the fitted curve to the $\xi(s)$ for the main sample, plotted here as a reference for comparison. }
\label{xi_3subz}
\end{center}
\end{figure*}

\begin{table}
\begin{tabular}{ccccc}
\hline
\hline
$\Delta z$ & $\bar z$ & \# of & \# of \textsc{ngc}& n$~(10^{-6})$\\
 & & quasars &  quasars & $(h^{-1}\rm Mpc)^{-3}$\\ \hline

$2.20 \le z < 2.39$ & 2.297  & 30824 & 24667 & $4.283\pm0.027 $\\
$2.39 \le z < 2.64$ & 2.497  & 30707 & 24493 & $3.001\pm0.019 $\\
$2.64 \le z\le 3.40$ & 2.971  & 30816 & 24724 & $1.057\pm0.007 $\\
\hline
\end{tabular}

\caption{Properties of the three \textsc{core} subsamples over the redshift
range $2.2 \le z \le 3.4$. The columns are redshift range, average redshift and
total number
of quasars for each redshift subsample with or without \textsc{sgc} quasars.
 The final column is the number density of the NGC quasars in each subsample, together with Poisson errors.}

\label{tabz}
\end{table}

\subsection{Redshift dependence}\label{relev}
In order to have sufficient dynamic range to characterize 
the evolution of quasar clustering, we extend the redshift upper limit we study
from 2.8 to 3.4. We divide the resulting sample of 73{,}884 \textsc{ngc-core} 
quasars\footnote{All quasars still have z\textsc{warning}=0 and completeness
greater than 75\%.} into three subsamples and measure the projected 2PCF both in
real and redshift space for each subsample. Table \ref{tabz} summarizes the
properties of the three redshift subsamples. The dividing redshifts (2.384 and
2.643) were selected such that the three subsamples contain almost the same
number of quasars.

Fig.\ \ref{wp_3subz} shows the result of measuring the real-space 2PCF as a function of redshift.
Motivated by the fitting result of the \textsc{ngc-core} sample described in $\S$\ref{realACF}, 
we fit a power law of index $\gamma=2$ over scales of $4 \lesssim r_{\rm p} 
\lesssim 25\,h^{-1} \rm Mpc$ for each of the three redshift bins. The scatter of the 
points (10 points are included in the fit for each 2PCF) results in higher uncertainties 
on the best-fit correlation length---but $r_0$ remains similar as a function of redshift. 

To evaluate how much fixing the power-law index affects our results, we allow
$\gamma$ 
to float as a second free parameter in the fit over the same range. There is a
considerable 
difference between the assumed $\gamma=2.0$ and the best-fit value for some
redshift bins ($\gamma_{1}=1.57 \pm 0.12$, $\gamma_{2}=1.69\pm 0.37$ and
$\gamma_{3}=1.88\pm0.29$). However, the best-fit values for $r_{0}$
remain consistent within the error bars.

\begin{figure*}
\begin{center}
\includegraphics[angle=0,scale=0.63]{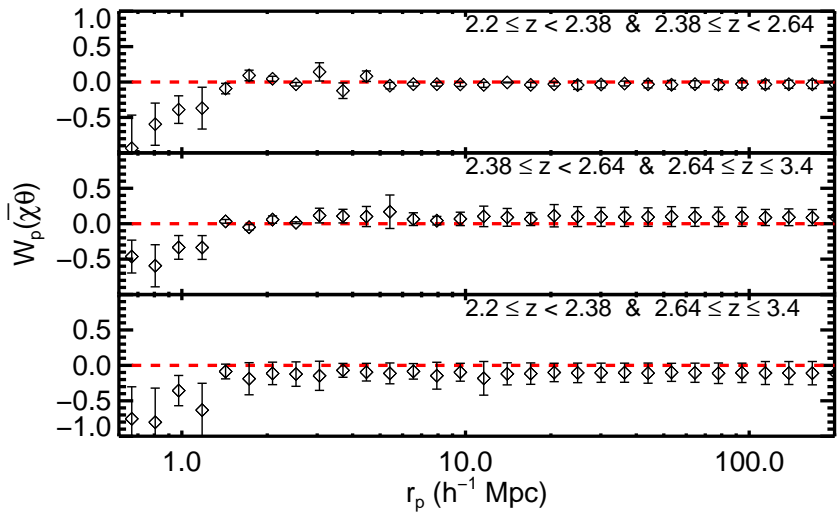}
\caption{The one-dimensional projected cross-correlation function
$\rm w_{p}(\bar \chi \theta)$ between quasars in different redshift subsamples. 
Lack of any strong correlation between redshift subsamples 
demonstrates that our masking procedure and pipeline-based redshift 
determinations are robust over our fitting scales.
We measure only the coordinate transverse to the line-of-sight
($r_{\rm p}$), as different
redshift bins are correlated along the line-of-sight. The data deviate from the from
the dashed line
(zero correlation) for scales below $\sim1.5\,h^{-1} \rm Mpc$; we discard these points for
all of the fits. 
This deviation is most likely caused by the fact that \textsc{boss} can not
place fibers at separations smaller than 62\arcsec\ ($1.3\,h^{-1} \rm Mpc$ at $z=2.5$)
on a single plate. 
We note that the slight deviations ($< \pm 0.1$) from zero at scales of $> 1.5 
\hmpc$ in the lower two panels are insignificant given the high covariance of 
the data on different scales. Nevertheless, these deviations could indicate 
low-level, correlated substructure in our masks that might, e.g., depend on how 
the depth of the quasar target sample changes with redshift.}
\label{subz_cross}
\end{center}
\end{figure*}

\begin{table*}
\begin{center}
\begin{tabular}{cccccccc}
\hline
\hline
$\Delta z$ &  $\Delta M_{i}$ & $\bar z$ & \# of quasars  & $\rm b_{Q}$ & $\chi^{2}_{\rm red}$ & $\rm r_{0}$($h^{-1} \rm Mpc$) & $\chi^{2}_{\rm red}$ \\

\hline
$2.20 \le z\le  2.80$ & $-28.74 \le M_{i} \le -23.78 $ & 2.434 & 55826 & 3.54 $\pm$ 0.10 & 1.06 &  8.12 $\pm$ 0.22 & 0.25 \\
\hline
$2.20 \le z < 2.384$ &  $-28.70 \le M_{i} \le -23.95 $& 2.297 & 24667& 3.69 $\pm$ 0.11 & 1.55 & 8.68 $\pm$ 0.35 &  0.45 \\
$2.384 \le z < 2.643$ & $ -28.74 \le M_{i} \le -24.11$&  2.497 & 24493 & 3.55 $\pm$ 0.15 & 0.61 & 8.42 $\pm$ 0.54 & 0.26 \\
$2.643 \le z\le 3.40$ & $-29.31 \le M_{i} \le -24.40 $& 2.971 & 24724 & 3.57 $\pm$ 0.09 & 0.66 & 7.59 $\pm$ 0.66 &  0.33\\
\hline
$2.20 \le z\le  2.80$ & $-28.74 \le M_{i} < -26.19$ & 2.456  & 18477 & 3.69 $\pm$ 0.10 & 2.19 & 8.62 $\pm$ 0.27 & 0.94 \\
$2.20 \le z\le  2.80$ & $-26.19\le M_{i} <  -25.36$ & 2.436 & 18790 & 3.56 $\pm$ 0.13 & 1.71 & 7.94 $\pm$ 0.41 & 1.63 \\
$2.20 \le z\le  2.80$ & $-25.36\le M_{i} \le -23.78$ & 2.411 & 18559 & 3.81 $\pm$ 0.19 & 0.4 & 8.29 $\pm$ 0.36 & 3.29 \\
\hline

\end{tabular}
\end{center}

\caption{Clustering results for \textsc{ngc-core} sample and subsamples. The first five columns are redshift and absolute magnitude range,
their average values and the total number of NGC quasars in each of the luminosity
and redshift subsamples. Columns 6, 7 and 8 are the bias values that best fit
our measured redshift-space correlation function, their uncertainty and the
reduced $\chi^{2}$ (over 7 degrees of freedom for the main sample and 9 degrees of freedom for the redshift and luminosity
subsamples). Column 9 is the best
fit correlation length which is the result of fitting a one parameter power-law
with a fixed index of 2.0 to the measured real-space correlation function of
quasars in each subsample. The last two columns are the uncertainty on the
correlation length and its reduced $\chi^{2}$ (over 8 degrees of freedom for the redshift and luminosity
subsamples and over 9 degrees of freedom for the main sample). The lower limits on $\Delta M_i$ for the
redshift subsamples and full samples are the highest luminosity
quasars that happen to occur in the sample and no imposed upper limit in the selection is applied.}
\label{tabrb}
\end{table*}

We measure the redshift-space correlation function for the same three redshift-based 
subsamples. Similar to our approach for the real-space correlation function, we fit 
$\xi(s)$ for the three subsamples over the same fitting range as $\xi(\rm s)$ for the full \textsc{ngc-core} sample and measure the bias of quasars relative to dark matter in every bin of redshift. Fig.\ \ref{xi_3subz} displays $\xi(s)$ for the three subsamples. The model 
represented by the solid curve is as for Fig.\ \ref{xi}. Lower number densities for quasars in the subsamples produce noisier 
covariance matrices. As a result, we only use the diagonal elements of those matrices when fitting the real- and redshift-space correlation functions.

The upper part of Table \ref{tabrb} displays the fitting results for the evolution of the real 
and redshift-space correlation functions. Including full covariance 
matrices in our fits increases $\chi^{2}$ 
without producing a meaningful change in the best-fit value for the bias. The covariance matrices for $\xi$ are less noisy than for $w_{\rm p}$ on larger scales---but
extending the fits to larger scales for $\xi(s)$ (beyond the vertical dashed lines in Fig.\ \ref{xi} 
and Fig.\ \ref{xi_3subz}), still results in a large increase in uncertainty for the best-fit bias, presumably because the larger scale measurements are not well fit by the same bias factor and $\Lambda$CDM model that fits the data at $s \la 25 \hmpc$. 

As a further test of our overall masking procedures, we calculate the cross-correlation function
between quasars in different bins of redshift and display the results in Fig.\ \ref{subz_cross}.
The lack of any strong correlation between different redshift subsamples
demonstrates that the pipeline-based redshifts used for this study are of
sufficient reliability and that 
the masking procedure we use to produce the angular selection function for
\textsc{boss} quasars
is robust.  

\begin{figure*}
\begin{center}
\includegraphics[angle=0,scale=0.4]{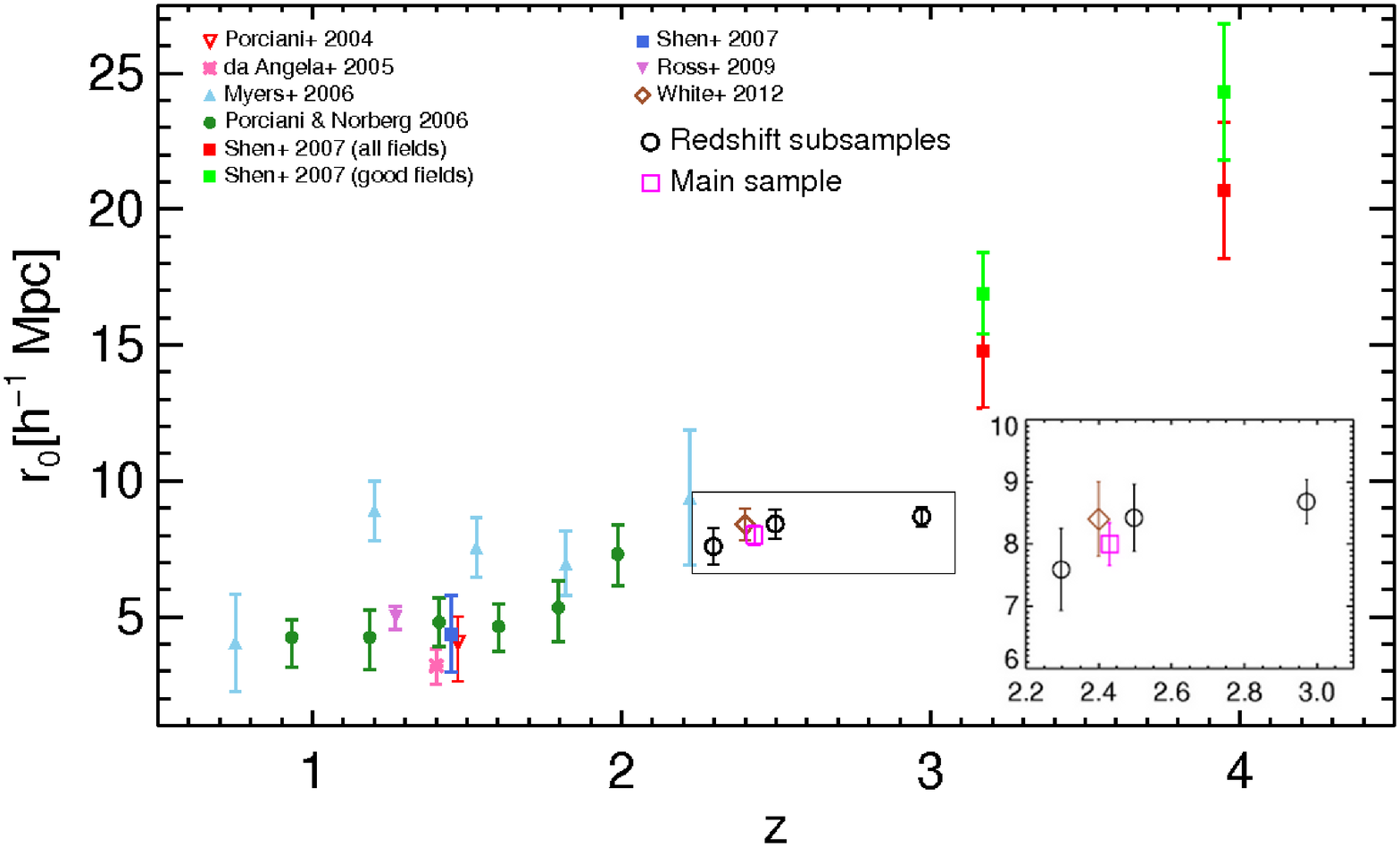}
\caption{Evolution of the correlation length $r_{\rm 0}$ with redshift. We display the results 
for all \textsc{ngc-core} quasars (open orange square) and for the different 
redshift ranges (open black circles) and absolute magnitude bins (filled black triangles)
studied in this work (see Table \ref{tabrb}). Results from prior works have been recalibrated 
to reflect our chosen cosmology. The inset provides a more detailed view of
the range $2.2 \leq z \leq 3.0$.}
\label{rzplot}
\end{center}
\end{figure*}

\begin{figure*}
\begin{center}
\includegraphics[angle=0,scale=0.4]{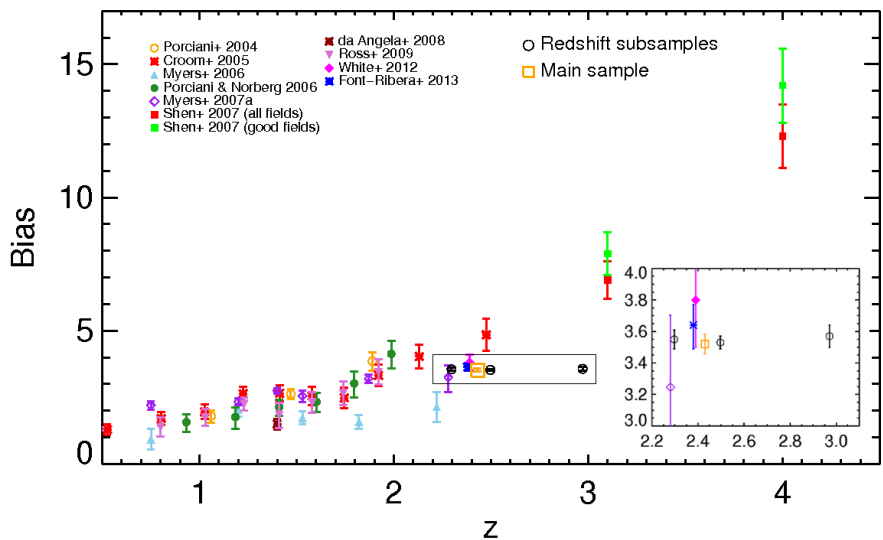}
\caption{Evolution of quasar bias with redshift. We display the results 
for all \textsc{ngc-core} quasars (open orange square) and for the different 
redshift ranges (open black circles) and absolute magnitude bins (filled black triangles)
studied in this work (see Table \ref{tabrb}). Results from prior works have been recalibrated 
to reflect our chosen cosmology. The inset provides a more detailed view of
the range $2.2 \leq z \leq 3.0$.}
\label{bzplot}
\end{center}
\end{figure*}

There have been few measurements of quasar clustering near $z\sim2.5$, 
which indicates the difficulties inherent in uniformly-selecting large numbers of quasars
near this redshift. 
Fig.\ \ref{rzplot} and Fig.\ \ref{bzplot} show the reported values of correlation length and 
quasar bias from previous works \citep{po04,cro05,da05,my06,po06,my07,she07,da08,ro09,wh12,fr13}. 

Some caution is required in interpreting Figures \ref{rzplot} and \ref{bzplot}, as
different samples probe different luminosity ranges and different
groups have made different assumptions and choices when inferring
$r_0$ and bias (e.g., power-law vs.\ $\Lambda$CDM fit, projected
or redshift-space fit, range of separations used) and when computing
the error bars on their results.  For Figure 13, we have corrected
all bias values to our adopted cosmology by multiplying the authors'
reported values and error bars by
$\sigma_{8,{\rm rep}}(z)/\sigma_{8,{\rm fid}}(z)$ where
$\sigma_{8,{\rm rep}}(z)$ is the value of $\sigma_8$ at
redshift $z$ in the reported cosmology and $\sigma_{8,{\rm fid}}(z)$
is the value for our fiducial cosmology (see \S 1).
\citet{she07} report values of $r_0$ for samples in
``all fields'' and ``good fields'' (with higher photometric
data quality but a smaller sample) at effective redshifts
$z = 3.1$ ($2.9 \leq z \leq 3.5$) and $z=4.0$ ($z \geq 3.5$).
For corresponding bias values in Figure \ref{bzplot} we use those calculated
by \citet{wh08} by fitting the Shen et al.\ measurements.

The BOSS measurements of quasar bias, from this paper and from the
Ly$\alpha$ forest cross-correlation measurement of
Font-Ribera et al.\ (2013), are the most precise constraints
at any redshift, and by far the most precise at $z > 2$.
Our value of $b_Q = 3.54 \pm 0.10$ for the redshift range
$2.2 \leq z \leq 2.8$ is compatible with Font-Ribera et al.'s
value of $3.64^{+0.13}_{-0.15}$ at $1\sigma$, a reassuring level of
consistency given the difference of measurement methods.
Our bias and $r_0$ values are higher than most of those measured
at $z < 2$, though these are typically for lower luminosity
thresholds so it is difficult to separate redshift and luminosity effects.
Our redshift subsample measurements are consistent with constant $r_0$
or constant bias over the redshift range $2.2 \leq z \leq 3.4$.
(Because matter clustering grows by a factor of 1.35 over this
redshift span, constancy of one quantity implies evolution of the other,
but our measurement errors are too large to definitively
establish evolution of either.)  The SDSS-based measurements
of Shen et al.\ (2007) are significantly higher than ours.
We discuss the 
  marked difference between our measurements and the Shen et al.\ (2007)
value at $z=3.1$ ($b_Q = 6.9 \pm 0.7$ for ``all fields'') 
in \S\ref{sec:HM}.

\begin{table}
\begin{tabular}{ccccc}
\hline
\hline
$ \Delta M_{i}$ & $\bar M_{i}$ & \# of & \# of \textsc{ngc}& n$~(10^{-6})$\\
 & & quasars &  quasars & $(h^{-1}\rm Mpc)^{-3}$\\

\hline
$[-28.74 , -26.19]$ & -26.87 & 23809  & 18477&  $0.982 \pm 0.0072$\\
$[-26.19 , -25.36]$ & -25.77 & 23669  & 18790 & $0.998 \pm 0.0073$\\
$[-25.36 , -23.78]$ &  -24.90 & 22499  & 18559& $0.986 \pm 0.0072$\\

\hline
\end{tabular}
\caption{Properties of the three \textsc{ngc-core} absolute magnitude subsamples over the redshift range $2.2 \le z \le 2.8$. The columns list the absolute magnitude range and its average for
each luminosity subsample as well as the total number of quasars in each
subsample with or without \textsc{sgc} quasars (3rd and 4th columns
respectively). $\bar M_{i}$ corresponds to the average absolute magnitudes of
the \textsc{ngc} quasars in each subsample.
 The last column is the number density and the Poisson errors of the quasars in 
each subsample.}

\label{tabmag}
\end{table}

\begin{figure*}
\begin{center}
\includegraphics[angle=0,scale=0.285]{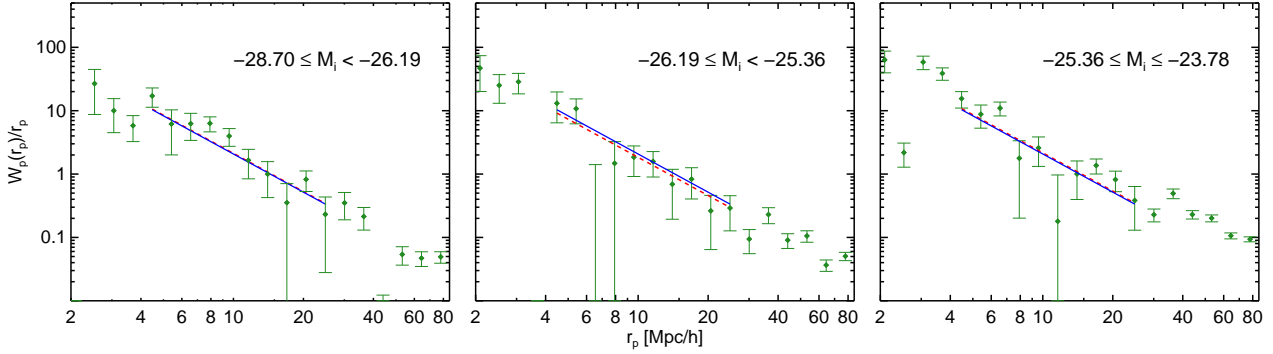}
\caption{To constrain any luminosity dependence of the clustering of
\textsc{boss} quasars, we divide the quasar sample into three absolute magnitude
subsamples covering the same redshift range ($2.2 \le z \le 2.8$). The three
panels display the real space-correlation function fitted by a one-parameter
power law over scales of $4 \lesssim r_{\rm p} \lesssim 25\,h^{-1} {\rm Mpc}$.
The red dashed line is the one-parameter fitted power-law with $\gamma=2.0$ and
the blue solid line is the one-parameter fitted power-law to the measured
correlation function of the main sample plotted here as a reference line.}
\label{wp_3subLUM}
\end{center}
\end{figure*}

\begin{figure*}
\begin{center}
\includegraphics[angle=0,scale=0.40]{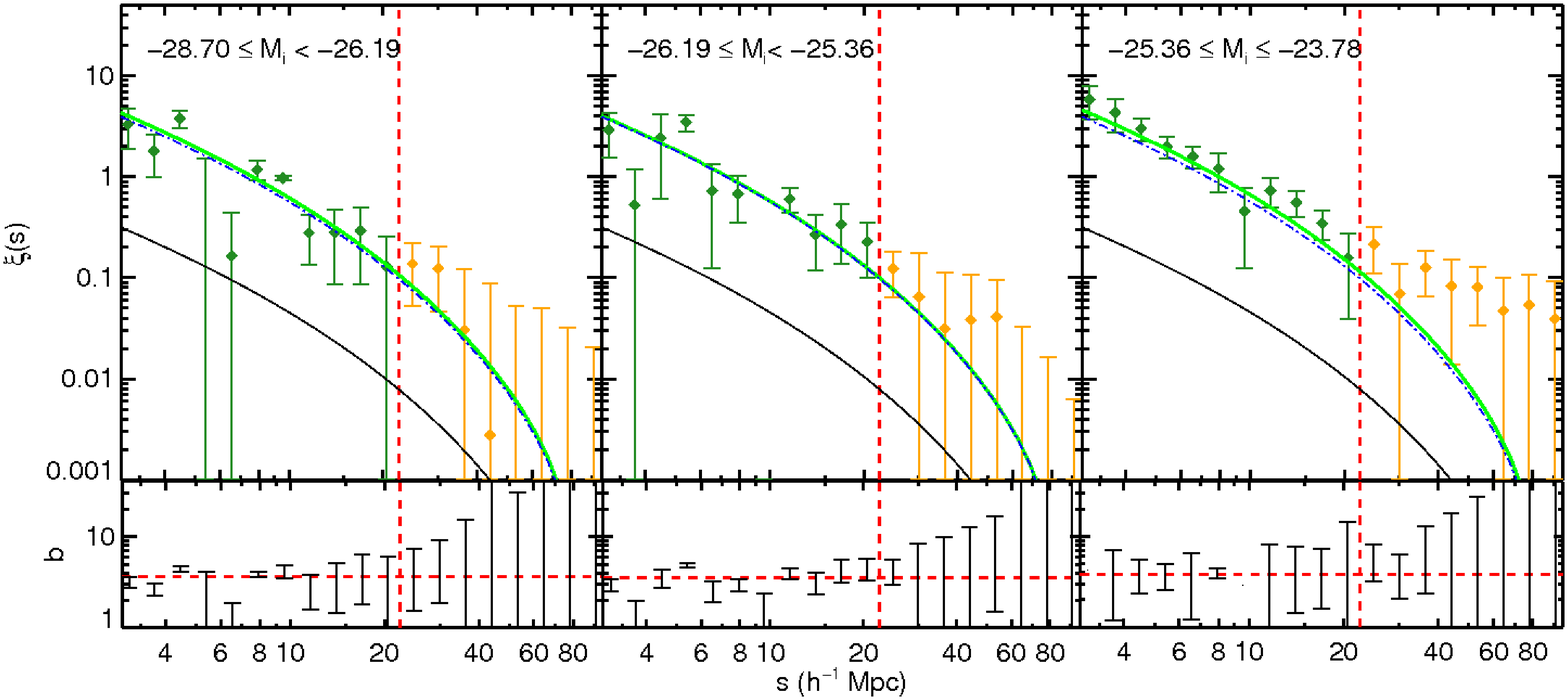}
\caption{Similar to Fig.\ \ref{xi_3subz} but for samples in the
redshift range $2.2 \le z \le 2.8$, subdivided by absolute magnitude. See Table \ref{tabrb}
 for the fitting results. The green curve is the fitted dark matter
correlation function to the measured $\xi(s)$ for each sample and the overlaid blue dashed curve
is the fitted curve to the $\xi(s)$ for the main sample, plotted here as a reference for comparison.}
\label{xi_3subLUM}
\end{center}
\end{figure*}

\subsection{Luminosity dependence}
\label{sec:lumdep1}

As described in \S\ref{sec:lumdata}, we divide the \textsc{ngc-core} sample of
55{,}826 quasars into three luminosity subsamples (see Table \ref{tabmag}), such
that there is a distinct 
difference between the average luminosity for each equally sized subsample. 
Fig.\ \ref{wp_3subLUM} and Fig.\ \ref{xi_3subLUM} present the results of measuring
the real and redshift-space correlation functions, respectively, for these
luminosity-based subsamples.  As the luminosity subsamples have even lower number densities than the redshift subsamples, the covariance 
matrices for the luminosity-based ACFs are noisier than those for the redshift
subsamples, so we
fit their real- and redshift-space correlation functions using only the diagonal elements of their covariance matrices. An assumption of diagonal error covariance may be fairly accurate
for these sparse samples, as shot noise makes a large contribution
to the error budget.

\begin{figure*}
\begin{center}
\includegraphics[angle=0,scale=0.42]{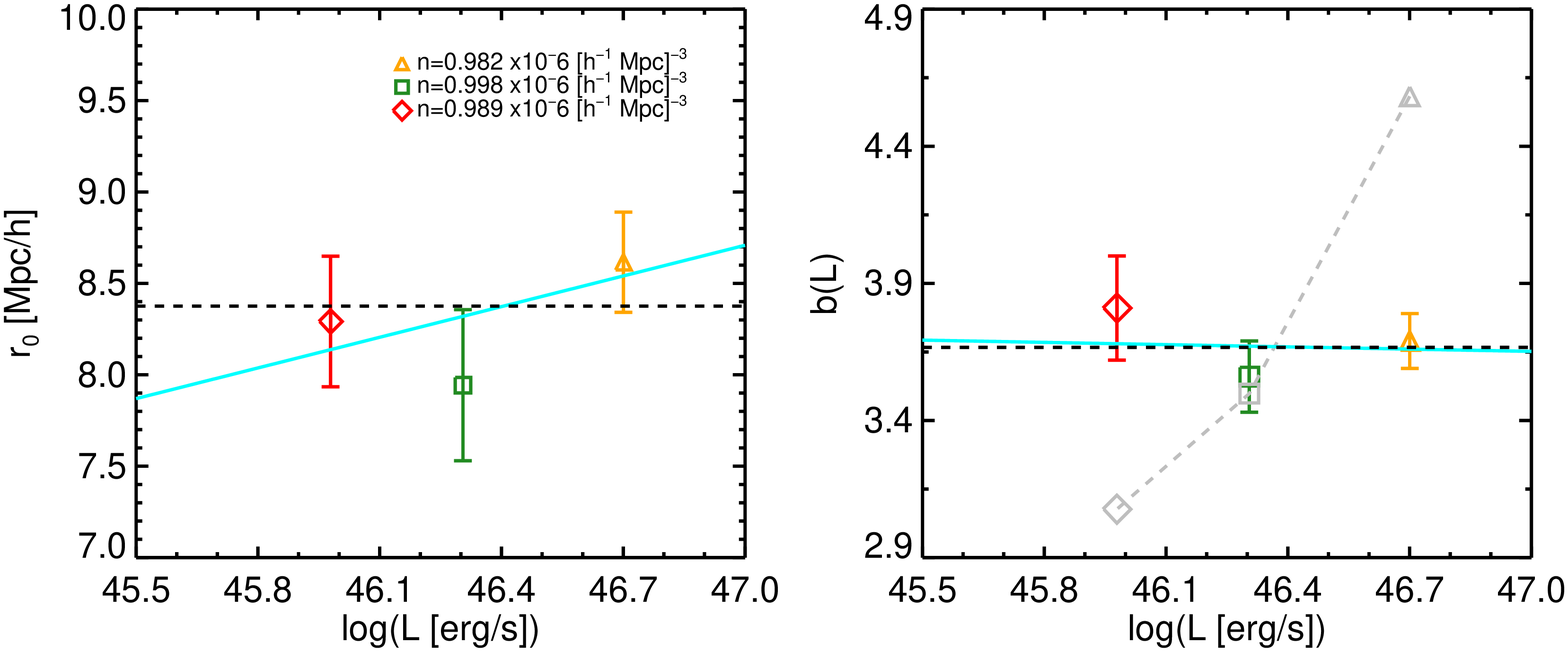}
\caption{The variation of the measured correlation length and measured and expected bias as a function of bolometric luminosity for quasars in three absolute magnitude subsamples. Each of the luminosity subsamples has a similar redshift ($z=2.456$, $z=2.436$ and $z=2.411$). The number density of quasars in each sample ($n$) is recorded in the legend of the left-hand panel. The blue solid line is the best-fit power law (in the general form of $a (L/L_{0})^{m}$ with $\rm log_{10}(L_{0})=46.33$) to the measured correlation length and bias values in the three luminosity bins. The black dashed line is the best-fit power law with $m=0$ (i.e. a model for which clustering does not depend on luminosity). 
In the right-hand panel, the coloured symbols are the best fit biases to the measured $\xi(s)$ for each sample. The slopes of the fitted line to the measured (coloured) biases and correlation lengths are $m = -0.027 \pm 0.27$ and $m=0.56\pm 0.80$ respectively. 
The connected grey points in the right hand panel show the predicted luminosity dependence of bias if
quasar luminosity is monotonically related to halo mass and all halos have the same quasar duty
cycle $f=0.0073$ that best fits the full sample.
}
\label{br0l}
\end{center}
\end{figure*}

Figure \ref{br0l} shows the dependence of $r_0$ and $b$ on luminosity for
these three subsamples, each with an effective redshift $z\approx 2.4.$
Because of the difference between fitting a power-law (for $r_0$) and
a $\Lambda$CDM correlation function (for $b$), the order of points
is not the same in the two panels, but it is clear that the data are
consistent with constant $r_0$ or $b$, independent of luminosity
over our measured range.  Solid curves show power-law fits of the
form $a (L/L_0)^m$ for which we find $m = 0.56 \pm 0.80$ (for $r_0$)
or $m = -0.027 \pm 0.27$ (for $b$), consistent with $m=0$.
Dashed horizontal lines show the best fit with $m$ forced to zero.
Our highest luminosity sample has a threshold
$M_i < -26.19$ vs.\ $M_i < -26.6$ for Shen et al.\ (2007).
The lack of either redshift or luminosity trends within our
sample therefore makes the significantly higher clustering of
their sample (spanning $2.9 \leq z \leq 3.5$) somewhat surprising.
However, we have not attempted to measure clustering for a
sample that mimics both their redshift range and luminosity threshold.

\section{Halo Masses and Duty Cycles}
\label{sec:HM}

Following the approach proposed in a range of previous works \citep[e.g.][]{ha01,mw01,wy05}, we
now use our clustering measurements to constrain the host halo
masses of BOSS quasars and the duty cycles of quasar activity
in these halos.  The constraint on halo mass comes from the
fact that more massive halos have higher clustering bias
\citep{kai84, tin10}, so the observed clustering
of quasars determines the characteristic mass of their host halos.
Moving from this halo mass to a duty cycle requires a further
assumption to relate luminosity and halo mass.
Here we assume that this relation is monotonic with no scatter---the black hole in a given halo is either ``on'' (i.e.\ observed as a quasar) or ``off"---and we test
whether this assumption is consistent with our observed (lack of)
luminosity dependence.

For each of our samples we compute
both a characteristic halo mass $\mhbar$ and a minimum halo
mass $\Mhmin$ (see Table~\ref{tabmass}). We calculate these
values using the formula of \citet{tin10} to model
$b(M)$. Specifically, we use their Eqn.\ 6 with
parameters chosen for an overdensity of $\Delta = 200$ and
 with our adopted cosmology. We
quote a characteristic mass as, simply, the mass that corresponds to our measured bias
in the \citet{tin10} formalism (i.e.\ $b_Q = b(\mhbar)$).  Alternatively, in our adopted model, $\Mhmin$ can be determined from the range of halos that correspond to a
given bias measurement via


\begin{equation}\label{eqn:bbar}
b(M > M_{h,{\rm min}}) \equiv \frac{\int_{M_{h,{\rm min}}}^{\infty} \frac{dn}{dM} b(M) dM} {\int_{M_{h,{\rm min}}}^{\infty} \frac{dn}{dM} dM}
\end{equation}

i.e.\ our measured bias $b_Q = b(M>\Mhmin)$ is then interpreted to be the mean bias of halos with mass above $\Mhmin$ weighted by
the halo abundance $dn/dM$, which we take from the \citet{tin08} halo mass function at the mean redshift appropriate to each of
our samples (see Table~\ref{tabrb}). 


Associating $\Mhmin$ with the minimum halo mass required to host a
quasar above the sample luminosity threshold implicitly assumes
that all halos above $\Mhmin$ have an equal probability of hosting
a sample quasar (i.e.\ constant duty cycle) and that halos below
$\Mhmin$ can only host quasars below the luminosity threshold (i.e.\ quasars
hosted in halos below a threshold of $\Mhmin$ are not present in our
sample).

\begin{figure*}
\begin{center}
\includegraphics[angle=0,scale=0.45]{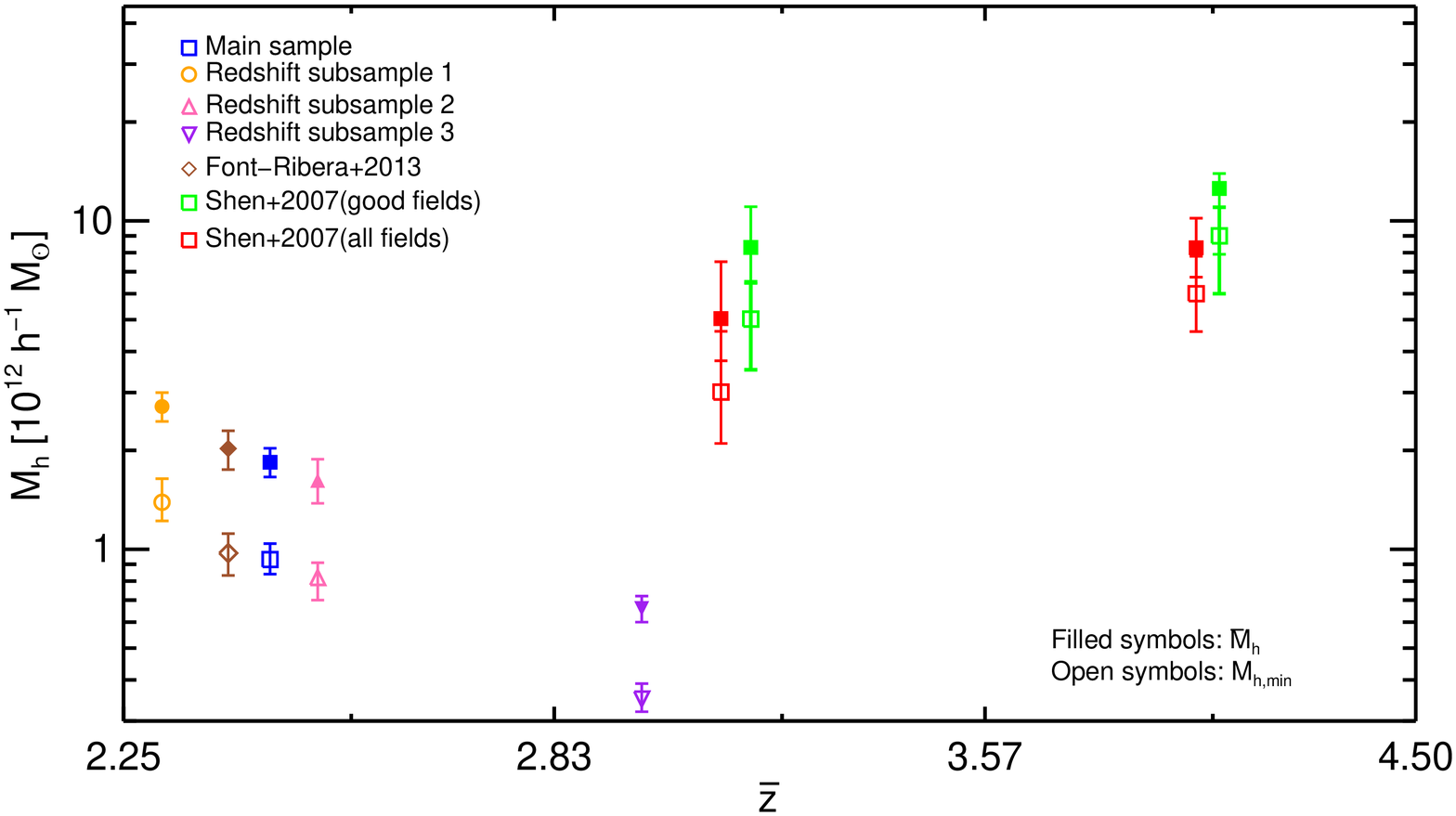}
\caption{The evolution of minimum and characteristic halo mass [defined respectively by $b_Q=b(M>\Mhmin)$
and $b_Q=b(\mhbar)$] for the main sample and three redshift subsamples (detailed in Table \ref{tabz}) using the halo mass function from \citet{tin10}. We also measure these quantities for \citet{she07} and \citet{fr13} based on their reported bias values. See Table \ref{tabmass} for the values and their 1-sigma uncertainties. The "good fields"  points have been offset slightly for visual clarity.}
\label{mzplot}
\end{center}
\end{figure*}

We determine $\mhbar$ and $\Mhmin$ and their errors by matching
the best-fit values and error range of $b_Q$ recorded in Table~\ref{tabrb}
 and
report these results in Table~\ref{tabmass}. We have also converted the \citet{fr13} measurements at $z=2.4$ and the 
\citet{she07} measurements
at $z=3.1$ and $z=4.0$ to halo masses using the same formalism and
display these alongside our results in Figure \ref{mzplot}. Unsurprisingly, our halo mass results are in excellent
agreement with \citet{fr13}, which are also drawn from the {\sc boss} survey and which therefore probe a similar
sample of quasars to that used in our analysis. It is more difficult to bring our results into accord with \citet{she07}, who measure 
host halo masses for $z\sim3.1$ quasars
that are an order of magnitude larger than our masses at $z\sim3.0$. The observational origin of the larger
measured masses in \citet{she07} is clear from Fig.\ \ref{rzplot} and Fig.\ \ref{bzplot}. Our clustering bias remains flat with
redshift and as characteristic halo masses are smaller earlier in cosmic history, our flat $b(z)$ implies a dwindling halo mass 
at higher redshift. The biases measured by \citet{she07}, on the other hand, 
greatly exceed our measurements at earlier times, implying that $b(z)$ grows steeply with redshift and that
quasars would therefore occupy higher-mass halos at earlier times. 

 The higher biases measured by \citet{she07} ultimately
have their root in the raw clustering measurements displayed in Fig.\ \ref{xicomp}. Much of the difference between our measurement in Fig.\ \ref{xicomp} and that of
\citet{she07} is driven by data on large scales ($s > 30\,h^{-1} {\rm Mpc}$), and, in particular,
by a single bin around $\sim 35 \,h^{-1} {\rm Mpc}$.
\citet{she07} fit their clustering measurement to larger scales than we are willing to trust given our observational systematics ($s < 25\,h^{-1} {\rm Mpc}$), and it is possible that
the \citet{she07} results on larger scales are somewhat contaminated by similar systematics
that are obfuscated by the larger statistical errors associated with their measurement.
With this caveat in mind, however, we
will take the results of \citet{she07} at face value and assume that
their measured clustering on large scales is accurate. If
the high biases measured by \citet{she07} are correct, they would im-
ply a sharp change in host halo masses at luminosities or redshifts
slightly beyond the boundaries of the samples analyzed here.

To compute the duty cycles in Table~\ref{tabmass}, we compare the cumulative luminosity function
of quasars over a range of luminosities to the cumulative space density
of halos over a range of host halo masses \citep{ha01,mw01}:

\begin{equation}\label{eqn:fduty}
f_{\rm duty} = \frac{\int_{L_{\rm min}}^{L_{\rm max}} \Phi(L) dL}{\int_{M_{\rm h,min}}^{\infty} \frac{dn}{dM} dM}.
\end{equation}

We take the quasar luminosity function $\Phi(L)$ from \citet{ro13}, and report values of the number density based on this
luminosity function in Table~\ref{tabmass}. \citet{ro13} derive careful corrections to the number density of {\sc boss} quasars
by simulating quasars and attempting to reselect them with the {\sc boss} targeting algorithm. This work on the {\sc boss} selection
function implies that {\sc boss} only recovers $\sim65$\% of $z\sim2.5$ quasars at $i\sim20$ falling to only $\sim35\%$ of $z\sim2.5$ quasars at $i\sim21$.
\citep[see Table~4 of][]{ro13}.
This selection function is reflected in the differences in our characteristic number densities in Table~\ref{tabmag} and
the number densities derived from $\Phi(L)$ in Table~\ref{tabmass}. In concert, the values from these two tables imply that
our highest luminosity clustering sample is only $\sim55\%$ complete and that our lowest luminosity clustering sample is only $\sim30\%$ complete, in
good agreement with \citet{ro13}.

 Note that it is reasonable to integrate our halo masses to $\Mhmax=\infty$ in Eqn.\ \ref{eqn:fduty}, rather than to some other threshold value of $\Mhmax$. This is because, in the formalism of Eqn. \ \ref{eqn:fduty}, we are making no strong assumption that quasar clustering correlates strongly with quasar luminosity. In effect, we are assuming that quasars of a specific luminosity could occupy a broad range of halo masses. We relax this assumption later in this section by inverting our argument in order to predict the biases that would arise if halo mass was a tight, monotonic function of luminosity. In Table~\ref{tabmass} we show our results for $\fduty$, derived from Eqn.\ \ref{eqn:fduty}, for our main sample and for each of our redshift- and luminosity-based subsamples. Given the nearly constant values of $\Mhmin$ we derive from clustering, one can interpret the $\fduty$ values as the probability that a halo above $M_h \approx 10^{12}\,h^{-1} M_\odot$ hosts a quasar in the corresponding luminosity range, and these probabilities are approximately proportional to the space densities of each subsample.

\begin{figure}
\begin{center}
\includegraphics[angle=0,scale=0.30]{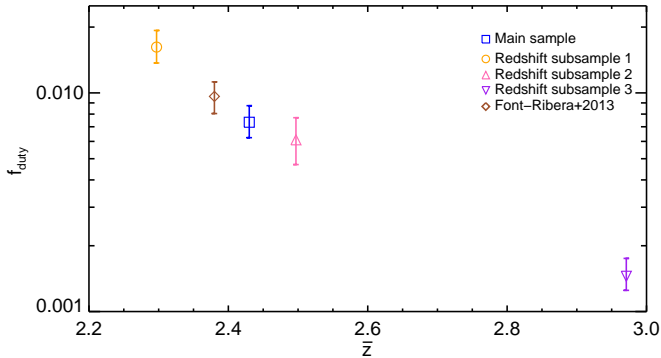}
\caption{The evolution of the duty cycle for the main sample and three redshift subsamples (detailed in Table \ref{tabz}) using the halo mass function from \citet{tin10} 
and the quasar luminosity function from \citet{ro13}. We also measure $\fduty$ for \citet{fr13}. See Table \ref{tabmass} for the $\fduty$ values and their 1$\sigma$ uncertainties.}
\label{dcplot}
\end{center}
\end{figure}

We derive error bars on $\fduty$ by matching the $\pm 1\sigma$
values of $\Mhmin$ inferred from the $b_Q$ uncertainties.
Figure \ref{dcplot} plots our derived values of $\fduty$ against redshift
for our primary sample and for our samples binned by redshift.
We do not plot $z<2$ values
because the implicit assumption of a monotonic relation between
luminosity and halo mass is likely to become inaccurate
at redshifts below $z \approx 2$ where the quasar luminosity
function begins to decline.
Our bias measurements imply a duty cycle of approximately 0.7\% for
our full sample, with a lower value of 0.1--0.2\% for
our highest redshift subsample. Disentangling whether this result is
driven by redshift or luminosity is difficult. By virtue of the magnitude-limited
nature of {\sc boss} our highest redshift subsample also contains moderately more
luminous quasars. 
It is plausible that the
most luminous quasars should
have smaller duty cycles.
Our results are reasonably
consistent with Fig.\ 14 of \citet{wh12}, which suggests that duty cycle is only
weakly dependent on luminosity at the level of 0.1--1\%.
However, we caution that the translation from quasar bias to duty
cycle relies on the assumption that the sharp threshold in quasar luminosity
corresponds to a sharp threshold in halo mass.

We can again compare our results to those of \citet{she07}, which imply higher duty cycles of $\sim 10$\%
at $z=3$ \citep{wh08,shan10a}. \citet{wh08} quote $b_Q\sim7$ or $b_Q\sim8$ 
for the \citet{she07} results using `all' data or just the `best' data. 
In our formalism, taking the \citet{she07} value of
$\Phi=5.592 \times 10^{-7}\,(h^{-1}\rm Mpc)^{-3}$ at $z\sim3$ (their Table 6) 
and adopting our mass function \citep{tin08}
and cosmology, we derive $f_{\rm duty} \sim 2$\% for
$b_Q\sim7$ and $f_{\rm duty} \sim 6$\% for $b_Q\sim8$. The factor of 3 difference
in these duty cycles is driven by the steep relationship between bias and mass as
halos become increasingly rare in the \citet{tin10} model. In contrast, we measure $f_{\rm duty} \sim 0.2$\% at $z\sim3$ for our sample.

The broad reason for this order-of-magnitude difference is that \citet{she07} measure far stronger clustering than we
measure, implying that their quasar sample is far more biased than our sample.
Their higher bias implies that their quasar sample occupies more massive and
therefore far rarer halos than our sample. Although their quasars are
brighter, and so less numerous than our quasars by a factor of 7, their measured bias
implies that their quasars are in halos that are a factor of $\sim70$ rarer than our halos (for their $b_Q\sim7$ result).
In other words, the lower density of the brighter \citet{she07} quasar sample compared to our
own does not compensate for the {\em far} lower density of the occupied halos implied by the
\citet{she07} measured bias. Taken at face value, 
the combination of our results with \citet{she07}
implies a sharp drop in the quasar duty cycle near $z \approx 3$. As we have previously
noted in this section, other possible reasons for this discrepancy might include the difficulty in measuring quasar
clustering for faint, high-redshift quasars, 
as well as uncertainties in the theoretical halo mass and bias function
at high redshift \citep{tin10}.

If quasar luminosity is a monotonic function of halo mass, then
more luminous quasars should be more strongly clustered at a given
redshift. As we have discussed previously in this section, this seems
unlikely given our measurement that bias is not a strong function of
luminosity. We can determine the biases that would be implied by
a monotonic relationship between halo mass and luminosity by
fixing $f_{\rm duty}$ in Eqn.\ \ref{eqn:fduty}. We can then assume that our
brightest quasar sample occupies a mass range of $\Mhmin < M < \infty$,
integrate over this range to determine $\Mhmin$ and then use this
$\Mhmin$ as the maximum halo mass $\Mhmax$ for our next most luminous sample.
In this manner, we can derive $M_h$ {\em ranges} for each of our
luminosity subsamples and then integrate over those ranges (from $\Mhmin$ to $\Mhmax$ rather than
always from $\Mhmin$ to $\infty$) in Eqn.\ \ref{eqn:bbar} to derive bias values.
 
To begin, we assume that the host halos for each of
our luminosity subsamples at $z=2.4$ have the value of
$\fduty = 0.0073$ that we derived for the full sample at this redshift.
For our highest luminosity bin, we then solve for the value of $\Mhmin$
from Eqn.\ \ref{eqn:fduty} setting $\Lmin = 2.99 \times 10^{46}\,{\rm erg\,s^{-1}}$
(see Table ~\ref{tabmass}) and $\fduty = 0.0073$, obtaining $\Mhmin = 2.39 \times 10^{12}h^{-1} M_\odot$.
For the next bin, we use Eqn.\ \ref{eqn:fduty} with $\fduty = 0.0073$ and
$\Lmin = 1.39\times 10^{46}\,{\rm erg\,s^{-1}}$, but now we set the
upper limits of the integrals to the {\it lower} limits
$\Lmin$ and $\Mhmin$ for the higher luminosity bin, deriving
a halo mass range
$\Mh/10^{12}h^{-1} M_\odot  = 1.39$--2.39 for this luminosity bin.
We repeat the process for the lowest luminosity bin, with
$\Lmax$ and $\Mhmax$ equal to the minimum values for the
middle luminosity bin, deriving a halo mass range
$\Mh/10^{12}h^{-1} M_\odot = 0.99$--1.39.

These calculations of $\Mhmin$, where we impose a duty cycle
and strictly require that more luminous quasars reside in higher
mass halos, are conceptually different from those reported in
Table \ref{tabmass}, where we derived $\Mhmin$ for each luminosity sample
independently based on its measured clustering bias.
Here we are deriving $\Mhmin$ and $\Mhmax$ from the luminosity
function, having chosen a duty cycle that will lead to the
correct average bias across the full sample.  Having derived $\Mhmin$ and $\Mhmax$
we then compute the predicted bias for each of our subsamples binned by luminosity
using Eqn.\ \ref{eqn:bbar}.



We show the three
predicted bias values as connected points in the right-hand
panel of Fig. \ref{br0l}.  
Previous quasar clustering measurements at lower redshifts have shown
little or no trend 
%
 with luminosity (e.g., \citealt{she09a,pad09}).  \citet{shan10b} found that these results required a tight relation between luminosity and mass for low values of the duty cycle $(\sim 0.0001)$ but quite considerable scatter in the mass-luminosity relation for higher values of the duty cycle $(\sim 0.001)$. In essence, the meaning of the duty cycle itself therefore becomes complex in the presence of substantial luminosity scatter (see also \citet{shan13}). Further, we should be cautious about the exact value of $f_{\rm duty}$ inferred for our full sample, which is used to derive the results in Fig. \ref{br0l}, because it is likely that {\em some} lower mass halos host {\em some} of the quasars in our sample. We must therefore be careful not to strictly interpret our results in Fig. \ref{br0l} as completely excluding a monotonic relationship between bias and luminosity. With these caveats in mind, however, the high value of $b_Q$ we have measured for quasar clustering near $z\sim2.5$ implies that we are on a steeply rising portion of the $b(M_h)$ relation, and the small statistical errors in Fig. \ref{br0l} suggest a near-constant value for $b(L)$ and a correspondingly constant $\Mhmin(L)$. Taken at face value, then, our results imply that all halos above $M_h\approx 10^{12}\,h^{-1}M_\odot$ are able to host a quasar anywhere in our sample luminosity range, with little correlation between luminosity and halo mass.

A number of more sophisticated approaches to quasar clustering are
possible, including abundance matching to galaxies \citep{co13},
halo occupation distribution (HOD) modeling \citep[e.g.][]{degraf11},
light curve models \citep{lid06}, models that add scatter
to the luminosity-halo mass relation \citep{shan10b},
and detailed evolutionary
models that track black hole growth and model their relation to halos
\citep[e.g.][]{shan10a}. We refer the reader to \citet{wh12} for a
more in-depth discussion of some of these issues.
The quasar bias factor, while a single number, provides a critical
quantitative constraint for any of these models.
We have now measured this constraint with high precision
at several luminosity thresholds over the redshift range $2.2 \leq z \leq 3.4$.


\begin{table*}
\begin{center}
\begin{tabular}{cccccc}
\hline
\hline
$\Delta z$ & $\Delta L$ & $\Phi(L_{\rm min}< L <L_{\rm max})$ & $\Mhmin$ & $\bar M_{\rm h}$  & $f_{\rm duty}$  \\
    &  ($10^{46}\,{\rm erg\,s^{-1}}$) & $(10^{-6}\,h^{-1}\rm Mpc)^{-3}$& \small ($10^{12}\,h^{-1} M_{\odot}$)& \small ($10^{12}\,h^{-1} M_{\odot}$) &  \\
\hline
$2.20 \le z \le 2.80$ &  $0.32 \le L \le 31.33 $  & $ 7.709^{+0.999}_{-1.109} $& $ 0.93^{+0.11}_{-0.09}$ & $1.84^{+0.19}_{-0.18}  $ & $0.00734^{+0.0014}_{-0.0011} $  \\

& \\

$2.20 \le z < 2.38$  & $ 0.38\le L \le 30.20 $ & $ 11.641^{+1.881}_{-1.806} $ &  $ 1.39^{+0.16}_{-0.15}$& $ 2.72^{+0.28}_{-0.27}$ & $0.0162^{+0.0031}_{-0.0025} $  \\

& \\
$2.38 \le z < 2.64$  & $ 0.44 \le L \le 31.33 $ & $ 7.804^{+1.693}_{-1.488} $  &  $ 0.82^{+0.15}_{-0.14} $& $ 1.62^{+0.26}_{-0.24} $ &  $0.0061^{+0.0016}_{-0.0014} $  \\

& \\
$2.64 \le z \le 3.40$  &$ 0.57 \le L \le 52.97  $ & $ 3.992^{+0.327}_{-0.565} $ &  $ 0.35^{+0.04}_{-0.07} $& $ 0.66^{+0.06}_{-0.06} $ & $0.00145^{+0.0003}_{-0.0002}$ \\

& \\
$2.20 \le z\le  2.80$  & $2.99 \le L \le 31.33$ & $1.846^{+0.137}_{-0.273} $  &  $ 1.03^{+0.12}_{-0.10}$ & $ 2.02^{+0.20}_{-0.19}$ &  $0.0019^{+0.0004}_{-0.0003}$ \\
& \\
$2.20 \le z\le  2.80$& $ 1.39 \le L < 2.99$ & $2.635^{+0.182}_{-0.468} $ & $ 0.92^{+0.12}_{-0.16}$ & $1.87^{+0.27}_{-0.24} $ &   $0.0023^{+0.0006}_{-0.0004}$ \\
& \\
$2.20 \le z\le  2.80$ & $ 0.32 \le L < 1.39$  & $3.228^{+0.955}_{-0.369} $  & $ 1.26^{+0.26}_{-0.15}$& $ 2.46 ^{+0.45}_{-0.40} $ &  $ 0.0044^{+0.0015}_{-0.0008}$  \\
\hline
\end{tabular}
\end{center}

\caption{The first three columns are the characteristics of each subsample;
the redshift range, the luminosity range and the number density of quasars calculated using the luminosity function of \citet{ro13}. The 4th and 5th columns are the minimum and the characteristic halo mass based on the halo mass function \citep{tin08} and bias model \citep{tin10}  at the average redshift of the sample (see Eqn.\ \ref{eqn:bbar}). The 6th column lists the 
duty cycle for the quasars in each sample, which is derived from $\Mhmin$ and $\Phi$ (see Eqn.\ \ref{eqn:fduty}). $f_{\rm duty}$ is expressed as a fraction of the Hubble time (9.785 $h^{-1}\,{\rm Gyrs}$). }
\label{tabmass}
\end{table*}

\section{Conclusions}
\label{sec:conclusions}

We measured the real and redshift-space correlation functions for a uniformly 
selected sample of 55{,}826 \textsc{ngc-core} quasars from the final data release of 
\textsc{sdss-iii/boss} over the redshift range $2.2 \le z \le 2.8$. We also
investigated the luminosity dependence of quasar clustering by splitting this
\textsc{ngc-core} sample into
three bins of absolute magnitude containing approximately equal numbers of
quasars. 
We do not detect a significant luminosity dependence to 
clustering strength for \textsc{boss} quasars over a factor $\approx 10$
range in luminosity.
Our clustering results are summarized in Table \ref{tabrb}.

This work provides precise quasar clustering measurements near $z\sim2.5$,
using samples of tens-of-thousands of quasars. 
We more than double the sample size compared to
the earlier {\sc boss} work of \citet{wh12}, and extend measurements of the 
real-space projected 2PCF to larger scales.
Our best-fit correlation lengths and slopes are in good agreement with, but
more accurate and more precise than, those of \citet{wh12}.

In order to investigate the redshift dependence of quasar clustering, we extended 
the upper redshift limit of our \textsc{boss} sample to $z=3.4$ and divided the 
resulting set of 73{,}884 quasars into three redshift subsamples of roughly equal size.
Our fitting results for the three redshift subsamples (see Table \ref{tabrb}), suggest that 
the correlation length does not evolve strongly over the redshift range $2.2 \le z \le 3.4$. 
When compared with earlier work (see Fig.\ \ref{rzplot}), this result suggests that $r_{0}$ 
evolves by no more than a factor $\sim 2$ 
from $z\sim 3$ to $z\sim 1$. We fit our redshift-space correlation
functions to a power spectrum for
dark matter halos from
{\sc halofit} \citep{sm03} modified by the correction
for redshift space distortions from Kaiser (1987). 
The bias of our quasars does not depend strongly on
either redshift or luminosity within our sample
and is roughly $b_Q\sim 3.5$ at $z\sim 2.5$.

We adopted the halo bias model of \citet{tin10} in order to use our measured biases to estimate the
average halo masses of quasars in our main (\textsc{ngc-core}) sample and in our three redshift subsamples
(see Table \ref{tabmass}). The estimated characteristic dark matter halo mass that we derived for quasars over the redshift range $2.2 \le
z \le 3.4$ covers a relatively wide range of halo masses from $\sim 0.6$ to $3 \times
10^{12}\,h^{-1} \rm M_{\odot}$ because a higher mass is required to yield the same bias
at a lower redshift.
By integrating our adopted bias model over the halo mass function of
\citet{tin08} we determined the minimum halo mass corresponding to our measured biases, deriving
$\Mhmin$ values from $\sim 0.3$ to $1.5 \times 10^{12}\,h^{-1} \rm M_{\odot}$. 

By comparing the number
density of halos above this mass threshold from \citet{tin08} to the quasar luminosity function for
{\sc boss} from \citet{ro13} we found that the duty cycles for our main and redshift subsamples are $f_{\rm duty} \sim 1$\%, with
our highest redshift (and therefore most luminous) subsample approaching $f_{\rm duty}\sim 0.2$\%.
These values of $f_{\rm duty} \sim 1$\% are in agreement with previous works, and are
broadly consistent with quasars being short-lived and episodic \citep[as argued by, e.g.,][]{cro05}
but are also not inconsistent with a mixed population of quasars---consisting of a bursty
population igniting as their parent halos grow, combined with a larger fraction of
shorter-lived quasars \citep[see, e.g., the discussion in][]{wh12}.
These values of $\fduty$ implicitly assume that halos with mass below $\Mhmin$ do not host quasars
in our sample, and they then describe the probability that a randomly selected halo above $\Mhmin$
is active above our sample luminosity threshold at any given time.

There are two somewhat surprising results in our analysis.  The first is a nearly constant
strength of clustering with luminosity.  We investigated a model in which the
quasar luminosity is a monotonic function of host halo mass, using the \citet{ross13}
luminosity function to infer halo mass ranges from quasar space densities assuming
the duty cycle ($\fduty \sim 0.7\%$) derived from fitting the full sample.
This model predicts a quasar bias that rises from $b_Q \sim 3.1$ for our faintest
sample to $b_Q\sim 4.5$ for our brightest sample, clearly inconsistent with our
measurements.  We conclude that there must be substantial scatter between halo
mass and quasar luminosity in this redshift/luminosity range.
 A broad short-timescale scatter in Eddington ratio has been proposed both 
from observational and theoretical standpoints in previous studies  
\citep[e.g.,][]{hop09,kauhec09,Nov11,Ai12,bon12,gab13,hic14,Vea14,sch15}, which 
further argues against a tight, monotonic relationship between luminosity and 
mass.
Our conclusion is stronger than that reached from previous studies at lower 
redshift
because of our high statistical precision and because the high bias of our quasars
puts us on the steeply rising portion of the $b(M_h)$ relation.

The second surprise is the difference between our measured host halo masses and duty cycles 
when compared to the work of \citet{she07}. At $z\sim3$ we derive halo masses and duty cycles that are close to
an order-of-magnitude smaller than those of
\citet{she07}. Both of these results are driven by \citet{she07} measuring that quasars cluster far more
strongly than we have found in this work. The high biases found by \citet{she07} imply that the quasars they use
are in far rarer halos than those we study, by a factor of $\sim70\times$. The quasars studied by \citet{she07}
are brighter than the {\sc boss} sample we study, and so they are also rarer, but only by a factor of $\sim7\times$. The fact
that the \citet{she07} quasar sample has a number density nearly $10\times$ lower than our {\sc boss} sample
but an implied halo abundance nearly $100\times$ lower is the main discrepancy between this earlier work and
our own results.
A strong dependence of quasar clustering on luminosity or redshift could explain the difference of
results, but no hints of such dependence are seen within our sample.

Perhaps the most likely explanation of these differences lies in the difficulty of studying
clustering of high redshift quasars targeted near the limits of an imaging survey.
This is particularly true near $z\sim 3$, where quasar colours resemble those of stars.
Our analysis takes advantage of improvements in SDSS photometric calibration and target
selection algorithms as well as much larger numbers that afford greater measurement
precision.  The excellent agreement between our measured quasar bias and the value
found by \citet{fr13} from cross-correlation with the Lyman-$\alpha$ forest is strong
evidence that our full sample result, at least, is not affected by variations in
target selection efficiency or redshift completeness (which would both affect
auto-correlations but not cross-correlations).  Nonetheless, the difference from
earlier results at $z>3$ highlights the importance of studying large quasar samples
selected from deep imaging with well controlled systematics.

This paper reports measurements of quasar clustering using the
final sample from the \textsc{sdss-iii/boss} survey. The prospects for further refining measurements of quasar clustering across a 
range of redshifts are excellent. The \textsc{sdss-iv}/e\textsc{boss} survey has just begun
(see Myers et al. 2015) and will ultimately result in a uniformly selected sample of 
$\sim$500{,}000 quasars over a range of redshift from $z\sim0.7$ to $z\sim3.5$.
\textsc{sdss-iv}/e\textsc{boss} should more than double the
number of uniformly selected quasars at moderate redshift ($2.2 < z < 3.5$) compared to the sample
used in this paper, and it will provide an opportunity to measure quasar clustering
across most of cosmic history in a single, consistently selected sample.

\section*{Acknowledgment}
Funding for SDSS-III\footnote{http://www.sdss3.org/} has been provided by the
Alfred P.\ Sloan Foundation, the Participating Institutions, the National
Science Foundation, and the U.S.\ Department of Energy Office of Science.
SDSS-III is managed by the Astrophysical Research Consortium for the
Participating Institutions of the SDSS-III Collaboration including the
University of Arizona, the Brazilian Participation Group, Brookhaven National
Laboratory, University of Cambridge, Carnegie Mellon University, University of
Florida, the French Participation Group, the German Participation Group, Harvard
University, the Instituto de Astroﬁsica de Canarias, the Michigan State/Notre
Dame/JINA Participation Group, Johns Hopkins University, Lawrence Berkeley
National Laboratory, Max Planck Institute for Astrophysics, Max Planck Institute
for Extraterrestrial Physics, New Mexico State University, New York University,
Ohio State University, Pennsylvania State University, University of Portsmouth,
Princeton University, the Spanish Participation Group, University of Tokyo,
University of Utah, Vanderbilt University, University of Virginia, University of
Washington, and Yale University.

SE and ADM were partially supported by NASA through ADAP award NNX12AE38G and
EPSCoR award NNX11AM18A and by the National Science Foundation through grant
number 1211112.

\appendix
\section{The ``BOSSQSOMASK'' software}

In order to make clustering measurements 
it is necessary to create a random catalog that mimics the selection function of the data
but that is otherwise unclustered (see $\S$\ref{dat}). Here we introduce the ``\textsc{bossqsomask}'' 
package---a series of reasonably mature IDL codes that can be used to produce random catalogs that mimic the {\sc boss} {\sc core}
quasar selection. \textsc{bossqsomask} has been used in previous papers 
\citep[e.g.,][]{wh12,kar14} but we make the current version available 
online\footnote{\url{http://faraday.uwyo.edu/~admyers/bossqsomask}} with the publication of this paper.  
Each \textsc{bossqsomask} routine has a header documenting its purpose, but
here we highlight the critical steps needed to produce coordinates and
redshifts for random objects according to the angular spectroscopic completeness 
and redshift distribution of \textsc{boss} quasars.

To mimic the selection of \textsc{boss} quasars, approximately 300,000 distinct spherical caps are created using the
\textsc{mangle}\footnote{{\sc mangle} is fully 
described and made freely available at \url{http://space.mit.edu/~molly/mangle/} }
software \citep{sw08}. To cull areas of the survey that greatly affect the
quasar density, we use veto masks developed to analyze \textsc{boss} galaxy 
clustering\footnote{\url{http://www.sdss3.org/dr10/tutorials/lss_galaxy.php}} 
\citep[see, e.g.,][]{wh11,an12,rossash12}. The angular completeness 
of the survey is tracked in regions defined by a unique set of overlapping spectroscopic tiles 
\citep[called a sector; see][]{bl03}. The angular completeness
in each sector (called $f_{\rm comp}$ in this paper) is defined to be the ratio of the 
number of quasars for which good spectra are obtained to the 
number of all targets in the target list. The \textsc{bossqsomask}
code takes the minimum acceptable
angular completeness as an input (e.g.,  $f_{\rm comp}$=0.75 can be sent to indicate 
that the survey should be limited to areas with angular completeness greater than 
75\%, as was used in this paper). 

In the rest of this appendix, we detail the specific
requirements and procedures for the \textsc{bossqsomask} software. There are numerous observational effects that have a second-order   effect on 
target density in \textsc{boss} \citep[e.g.,][]{rossash12}. Such effects are
particularly prevalent in the \textsc{sgc} region of \textsc{boss} (see the e\textsc{boss}
target selection papers, Myers et al., Prakash et al. 2015). \textsc{bossqsomask}
does not attempt to correct for these higher-order effects, and so users should not
trust the software to account for subtle differences in target density between the random
catalog and the quasar sample---differences that can cause deviations from zero clustering amplitude
on very large scales.

\par 
\subsection{File and system requirements} 
\textsc{bossqsomask} requires a list of all objects that would have been targeted as
a \textsc{boss} \textsc{core} quasar, i.e., all point sources in the 
\textsc{boss} area that have an {\sc xdqso} probability above 0.424 to the magnitude limit of \textsc{boss}. An appropriate 
file\footnote{e.g. the file {\tt  xdcore\_pqsomidzgt0.4\_targets.sweeps.fits} in the 
{\tt data/xdcore/} directory}
is included in the \textsc{bossqsomask} {\tt data} directory, and is used to make the 
psuedo-{\sc core} of all quasars that  ``should have been targeted'' (i.e. if angular
completeness was 100\%). 
\par

The package requires two environment variables to be defined\footnote{A detailed 
\textsc{readme} file guides the user through the steps of setting up the system to work with the package.}: \texttt{\$BOSSQSOMASK\_DIR} and \texttt{\$BOSSLSS\_DIR}. These variables are necessary
so that \textsc{bossqsomask} can find masks as well as code functions.
\texttt{\$BOSSLSS\_DIR} must contain all of the masking files listed as part of the galaxy
clustering package at, e.g., \url{http://www.sdss3.org/dr9/tutorials/lss_galaxy.php}.
A working copy of the \textsc{idlutils} software suite\footnote{e.g.\ \url{http://www.sdss3.org/dr8/software/idlutils.php}} is needed, and the binaries have to be properly compiled such that, 
e.g., \textsc{mangle}'s \textit{ransack} procedure in {\tt \$IDLUTILS\_DIR/src} is functional. 
Running the sequence of the following codes will produce a 
random catalog (with the user's desired size and completeness level) using 
the survey masks and list of \textsc{core} targets. 

\subsection{Main steps}

1) Making the \textsc{xd} \textsc{core} target list (\texttt{make\_xd\_core.pro}):\par
This code takes the file of all \textsc{xdqso} \textsc{core} targets 
as an input and constructs the pseudo-\textsc{core} 
target list (called {\tt data/xdcore/xdcore.fits}). The code uses veto masks 
for bright stars, center posts, bad $u$-columns and bad-photometry fields (many of 
these masks reside in the directory pointed to by \texttt{\$BOSSLSS\_DIR}).

\vspace{3 mm}
\noindent 2) Making the \textsc{xd} spAll file (\texttt{make\_xdspall.pro}):\par
This step requires the latest version of the \textsc{boss} {\tt spAll} 
file that contains the spectroscopic information for \textsc{boss} targets
(e.g. \texttt{spAll-v5\_7\_0}). The {\tt spAll} file can be
downloaded by running {\tt pro/get/get\_spall.pro}, which limits the 
{\tt spAll} file to just the relevant information (the ``{\tt mini-spAll}'' file)
and places it in the directory {\tt data/spall}.

Once the {\tt mini-spAll} file is in the correct path, \texttt{make\_xdspall.pro}
cross-matches observed objects 
in the {\tt spAll} file with targets from the {\tt xdcore.fits} file
(produced in the first step), and makes a file of matched targets
and spectra called {\tt data/compfiles/xdspall.fits}.

\vspace{3 mm}
\noindent 3) Assigning weights to the sectors of the survey mask (\texttt{make\_sector\_completeness.pro}):\par
In order to construct the random catalog with the correct angular density and to
select only sectors at the user-specified completeness level 
requires a weight that tracks the angular completeness for quasars in each sector of
 \textsc{boss}. This code takes the files of targets and spectra, obtains the \textsc{boss} sector
geometry from the \texttt{\$BOSSLSS\_DIR} and calculates the angular completeness
in each sector.



Several local files are written that are more manageable than the
comprehensive target files. Most notably the file {\tt data/compfiles/xdspallmask.fits},
which indicates which sector of the mask each target occupies. In addition,
smaller \textsc{mangle} polygon files are written that contain the subset of sectors that
are relevant to \textsc{boss} quasars ({\tt data/compfiles/bosspoly.fits} and 
{\tt data/compfiles/bosspoly.ply}).

\vspace{3 mm}
\noindent 4) Constructing the angular distribution of objects in the random catalog (\texttt{make\_angular\_random\_catalog.pro}):\par
This code reads the file of which sectors are relevant for \textsc{boss} quasars
(e.g., \texttt{bosspoly.ply}) as well as the
file that indicates the mask sector and weight for each
target (\texttt{xdspall.fits}) and creates files of targets and spectroscopically
confirmed targets based on the user-specified completeness (denoted {\em completeness}),
called {\tt xd\_comp{\em completeness}.fits} and {\tt xdspall\_comp{\em completeness}.fits}.

The code also determines the sectors that have a higher weight than the user-specified
completeness level, and then angularly populates those sectors (according to their weight)
with $N$ times more total random points than \textsc{boss} quasar
targets using \textsc{mangle}'s {\tt ransack}
procedure. $N$ is specified as an input by the user.

\vspace{3 mm}
\noindent 5) Assigning redshifts to objects in the random catalog (\texttt{make\_redshift\_random\_catalog.pro}): \par
Based on spectroscopically-confirmed quasars in the sectors derived 
by \texttt{make\_angular\_random\_catalog.pro}
this code generates a random redshift from the cumulative distribution of 
quasar redshifts. 
The code outputs final data and random catalogs in {\em fits} format of the size and 
at the completeness level specified by the user. As with all files generated during the construction
of the random catalog, the final output files are stored in the directory {\tt data/compfiles}.
The files are called {\tt data\_comp{\em completeness}.fits} and 
{\tt randoms\_comp{\em completeness}.fits} where completeness is the user-specified
completeness.

\subsection{Examining the produced random catalog}

Finally, the procedure \texttt{diagnostics.pro} is provided to help diagnose whether 
\textsc{bossqsomask} is functioning as expected. \texttt{diagnostics.pro} 
can conduct four different tests: 

\begin{enumerate}

\item \textit{Angular distribution test:} the angular distribution of data and random objects should match
\item \textit{Redshift distribution test:} the redshift distribution of data and random objects should match
\item \textit{Weights versus number of random points test:} the number of random points in each sector
should be proportional to ``completeness$\times$area'' for each sector at the user-specified completeness level
\item \textit{Weights versus number of objects test:} the number of data points in each sector 
should roughly be proportional to ``completeness$\times$area'' for each sector at 
the user-specified completeness level
\end{enumerate}



\bibliographystyle{mn2e.bst}
\bibliography{referencelist.bib}

\end{document}